\documentstyle[prl,aps,preprint,tighten,floats,aps,amssymb]{revtex}
\input epsf.tex

\newcommand{\bmat}{\left(\begin{array}}
\newcommand{\emat}{\end{array}\right)}
\newcommand{\beq}{\begin{equation}}
\newcommand{\eeq}{\end{equation}}

\newcommand{\drawsquare}[2]{\hbox{%
\rule{#2pt}{#1pt}\hskip-#2pt
\rule{#1pt}{#2pt}\hskip-#1pt
\rule[#1pt]{#1pt}{#2pt}}\rule[#1pt]{#2pt}{#2pt}\hskip-#2pt
\rule{#2pt}{#1pt}}

\newcommand{\fund}{\raisebox{-.5pt}{\drawsquare{6.5}{0.4}}}
\newcommand{\Ysymm}{\raisebox{-.5pt}{\drawsquare{6.5}{0.4}}\hskip-0.4pt%
        \raisebox{-.5pt}{\drawsquare{6.5}{0.4}}}
\newcommand{\Yasymm}{\raisebox{-3.5pt}{\drawsquare{6.5}{0.4}}\hskip-6.9pt%
        \raisebox{3pt}{\drawsquare{6.5}{0.4}}}
\newcommand{\antifund}{\overline{\fund}}

\def\NPB#1#2#3{Nucl. Phys. B {\bf #1} (19#2) #3}
\def\PLB#1#2#3{Phys. Lett. B {\bf #1} (19#2) #3}

\def\PRD#1#2#3{Phys. Rev. D {\bf #1} (19#2) #3}

\def\yzero{\smash{\hbox{$y\kern-4pt\raise1pt\hbox{${}^\circ$}$}}}

\def\-{\hphantom{-}}
\def\ov{\overline}
\def\s2{\frac{1}{\sqrt2}}

\def\beq{\begin{equation}}
\def\eeq{\end{equation}}
\def\beqa{\begin{eqnarray}}
\def\eeqa{\end{eqnarray}}

\def\diag{{\rm diag \,}}

\def\IF{\relax{\rm I\kern-.18em F}}
\def\II{\relax{\rm I\kern-.18em I}}
\def\IP{\relax{\rm I\kern-.18em P}}

\def\Dsl{\,\raise.15ex\hbox{/}\mkern-13.5mu D} 

\def\IC{\bf C}
\def\IZ{\bf Z}
\def\IT{\bf T}
\def\z2z2{$\IC^3/(\IZ_2\times\IZ_2)$}

\def\id{{\bf 1}}

\def\s{\sigma}

\def\z{\zeta}

\def\bo{{\raise-.3ex\hbox{\large$\Box$}}}               
\def\face{{\raise.2ex\hbox{$\displaystyle \bigodot$}\mskip-2.2mu \llap {$\ddot
        \smile$}}}                                      

\def\leftrightarrowfill{$\mathsurround=0pt \mathord\leftarrow \mkern-6mu
        \cleaders\hbox{$\mkern-2mu \mathord- \mkern-2mu$}\hfill
        \mkern-6mu \mathord\rightarrow$}       
\def\dvec#1{\vbox{\ialign{##\crcr
        \leftrightarrowfill\crcr\noalign{\kern-1pt\nointerlineskip}
        $\hfil\displaystyle{#1}\hfil$\crcr}}}           

\def\beq{\begin{equation}}
\def\eeq{\end{equation}}

\def\beqx{\begin{displaymath}}
\def\eeqx{\end{displaymath}}

\def\beqa{\begin{eqnarray}}
\def\eeqa{\end{eqnarray}}

\begin{document}
\draft
\date{\today}
\title{
\normalsize
\mbox{ }\hspace{\fill}
\begin{minipage}{12 cm}
RUNHETC-2002-47,
UPR-1023-T,
PUPT-2068,
MAD-TH-02-4 \\
{\tt hep-th/0212177}{\hfill}
\end{minipage}\\[5ex]
{\large\bf Supersymmetric Three Family SU(5) Grand Unified Models
from  Type IIA Orientifolds
with   Intersecting D6-Branes
\\[1ex]}}

\author{ Mirjam Cveti\v c$^{1}$\footnote{On Sabbatic Leave from the University of Pennsylvania},  Ioannis Papadimitriou$^{2}$\footnote{Exchange Scholar from the University of Pennsylvania}, and
Gary Shiu$^{3}$ }
\address{$^1$ Department of Physics and Astronomy,\\ Rutgers University,
Piscataway, NJ 08855-0849, USA\\
$^2$Physics Department, Princeton University, \\
Princeton NJ 08544, USA\\
  $^3$Department of Physics,
  \\University of Wisconsin, Madison, WI 53706, USA}

\maketitle

\thispagestyle{empty}

\vspace*{-1cm}

\begin{abstract}
We   construct   some $N=1$ supersymmetric  three-family SU(5)
Grand Unified Models
from type IIA orientifolds  on $\IT^6/(\IZ_2\times \IZ_2)$ with D6-branes
intersecting at general angles. These constructions are supersymmetric only for special
choices of untwisted moduli.
  We show that within the  above class of constructions  there are no  supersymmetric
  three-family models  with 3 copies of {\bf 10}-plets unless there are
simultaneously some {\bf 15}-plets.
We
systematically
analyze the construction of such models and their spectra.
The M-theory lifts  of these brane constructions
become purely geometrical backgrounds:
they are singular $G_2$ manifolds
where the Grand Unified gauge symmetries and three families
of chiral  fermions are localized at codimension 4 and codimension 7
singularities respectively.
We  also study some
preliminary phenomenological features of the models.
\end{abstract}
\newpage

\section{Introduction}
Grand unification \cite{GUT}
is an attractive possibility
of physics beyond the Standard Model, as it provides a natural explanation for
the unification  of strong and
electro-weak forces at
an energy  scale of the order of
$10^{15-16}$ GeV.  Over the years, many
Grand Unified Theories (GUTs) have been
proposed and their phenomenological features have been
thoroughly analyzed.
The fact that the GUT scale is remarkably close to the Planck (or string) scale is
tantalizing as it suggests that in addition to the elementary particle forces, unification may
include also naturally gravity.
It is therefore a relevant question as to whether grand unification is realized in
string theory, and if so, in what way does the GUT symmetry arise.

It is difficult to address these questions without some concrete models at hand.
To make progress, it is important to develop techniques
of constructing GUT models (as well as other extensions of the Standard
Model)
from string theory.
With the insights gained from studying
some concrete models, we can then
examine
whether string theory could shed new light on some of
the long-standing problems in
grand unification.
Furthermore, we can contrast these
constructions with string models that exhibit the Standard Model gauge symmetry
at the
string scale to understand what are the advantages and shortcomings of GUTs
in the framework of string theory.

These issues surrounding grand unification have been explored  extensively
in the context of weakly coupled heterotic string \cite{StringGUTreview}.
In recent years, however, the emergence of M theory has opened up many new avenues
for the construction of consistent string models.
In particular,
the advent of D-branes has allowed us to construct
 open string models that are non-perturbative from
the dual
heterotic string description \cite{PolchinskiWitten}.
The techniques of conformal field theory in describing D-branes
and orientifold planes in exactly solvable backgrounds (especially
orbifolds) have played a key role in the construction of
 four-dimensional chiral
models with~${\cal N}=1$ supersymmetry.
There are two broad ways in which chiral theories can be constructed from D-branes.
In the Type II orientifold models of
Refs. \cite{ABPSS,berkooz,N1orientifolds,zwart,ShiuTye,lpt,afiv,wlm,CPW,kr,CUW,aiqu},
chiral fermions appear on the worldvolume of D-branes when the branes are
located at 
singularities.
In this context,  an example of a three-family $SU(5)$
GUT model was constructed in \cite{lpt}.
However, in this model
there are no Higgs fields either
to break the $SU(5)$ gauge symmetry or to give rise to the
$SU(2)$ Higgs doublets of the Standard Model. Hence the model is
not fully realistic for futher phenomenological studies.

Another context
in which chiral fermions arise is when D-branes intersect at
angles \cite{bdl}. The spectrum of open strings stretched between
the intersecting D-branes contains chiral fermions which are
localized at the intersection. This fact was employed in
\cite{bgkl,afiru,bkl,imr,magnetised} (and subsequently in
\cite{bonn,bklo,bailin,kokorelis}) in the construction of non-supersymmetric
brane world models. In particular, numerous examples of
three-family Standard-like models as well as GUT models were
obtained. However, the dynamics to determine the stability of
non-supersymmetric models are not well understood, especially when
the string scale is close to the Planck scale (since the
non-supersymmetric models  are subject to large quantum
corrections). Typically, the models are unstable when D-branes are
intersecting at angles (since supersymmetry is generically broken).
Nevertheless, supersymmetric orientifold models with branes at
angles have been constructed \cite{CSU1,CSU2,CSU3},  resulting 
in the first examples of ${\cal N}=1$ supersymmetric
four-dimensional models with the quasi-realistic features of the
Standard Model in this context.  
Subesequently, the phenomenological features of
this class of models were explored in \cite{CLS1,CLS2,CLW}.   In
addition to the  Standard-like Models,
 an example of a superymmetric $SU(5)$ GUT model
with four families  of quarks and leptons
  (i.e., a net number of four ${\bf 10}$-plets and four ${\bf {\bar 5}}$-plets) was
presented in \cite{CSU2}.
The purpose of this paper is to extend this analysis and
further
explore the possibilities of constructing
more realistic supersymmetric $SU(5)$ models in this framework \footnote{Recently, a supersymmetric three-family  left-right
symmetric model based on ${\bf T^6/Z_4}$ orientifold was constructed \cite{blumrecent}. }.

Just like the models in \cite{CSU1,CSU2,CSU3}, the supersymmetric
orientifold models considered here correspond in the strong
coupling limit to compactifications of M theory on certain
singular $G_2$ manifolds. As discussed in \cite{CSU3}, the D-brane
picture provides a simple desciption of how chiral fermions arise
from singularities of $G_2$ compactifications
\cite{AW,Witten,aW,CSU1,CSU2}. More recently, there have been some
interests in exploring the phenomenological properties
(e.g., the problem of doublet-triplet splitting, threshold
corrections, and proton decay) of GUT models derived from $G_2$
compactifications \cite{G2-23-split,Friedmann}. It is therefore
interesting to explore if the features suggested in
\cite{G2-23-split,Friedmann} apply to this class of orientifold
models.

The purpose of this paper is few-fold. We shall systematize the
techniques of orientifold constructions with intersecting branes
to facilate the search for realistic models.
We consider
the most general intersecting D6-brane configurations
that are compatible with supersymmetry.
We then perform a systematic search for three-family
SU(5) GUT models within this framework for the case of
$T^6/{\bf Z}_2\times {\bf Z}_2$, and
show that in this construction there are no three-family models
(i.e., models containing three copies of ${\bf 10}$-plets) unless
some ${\bf 15}$-plets are present.
We therefore relax our criteria by allowing for the appearance of ${\bf 15}$-plets and
systematically construct some three-family GUT models. We also
briefly explore
the phenomenological features of these constructions.

This paper is organized as follows. In Section 2 we briefly
summarize the constraints in constructing supersymmetric
orientifold models with branes at angles. We have changed our
notation slightly from \cite{CSU1,CSU2} in order to simplify the
consistency conditions (tadpole cancellations) and supersymmetry
constraints. In Section 3, we classify the brane configurations
that preserve supersymmetry. In Section 4, we discuss in detail
how to perform a systematic search for the three-family $SU(5)$
models. The spectra of some of these models are tabulated in the
appendix. In Section 5 we conclude and briefly discuss some
physics implications as well as potential phenomenological
difficulties of these models.

\section{$\IZ_2\times \IZ_2$ orientifold models with branes at angles}

\label{rules} The rules to construct supersymmetric type IIA
orientifolds on $\IT^6/(\IZ_2\times \IZ_2)$ with D6-branes at
generic angles, and to obtain the spectrum of massless states were
discussed in \cite{CSU2}. In this section we recall the essential
points of the construction and emphasize some changes in notation
which could greatly simplify the systematic search for consistent
models.

We start with type IIA theory on $\IT^6/(\IZ_2\times
\IZ_2)$, where the orbifold group generators $\theta$, $\omega$  act on the
complexified coordinates on $\IT^6$ as \beqa
& \theta: & (z_1,z_2,z_3) \to (-z_1,-z_2,z_3) \nonumber \\
& \omega: & (z_1,z_2,z_3) \to (z_1,-z_2,-z_3). \eeqa We assume
$\IT^6$ can
be written as a product of three two-tori.

\begin{figure}
\begin{center}
\centering
\epsfysize=5.5cm
\leavevmode
\epsfbox{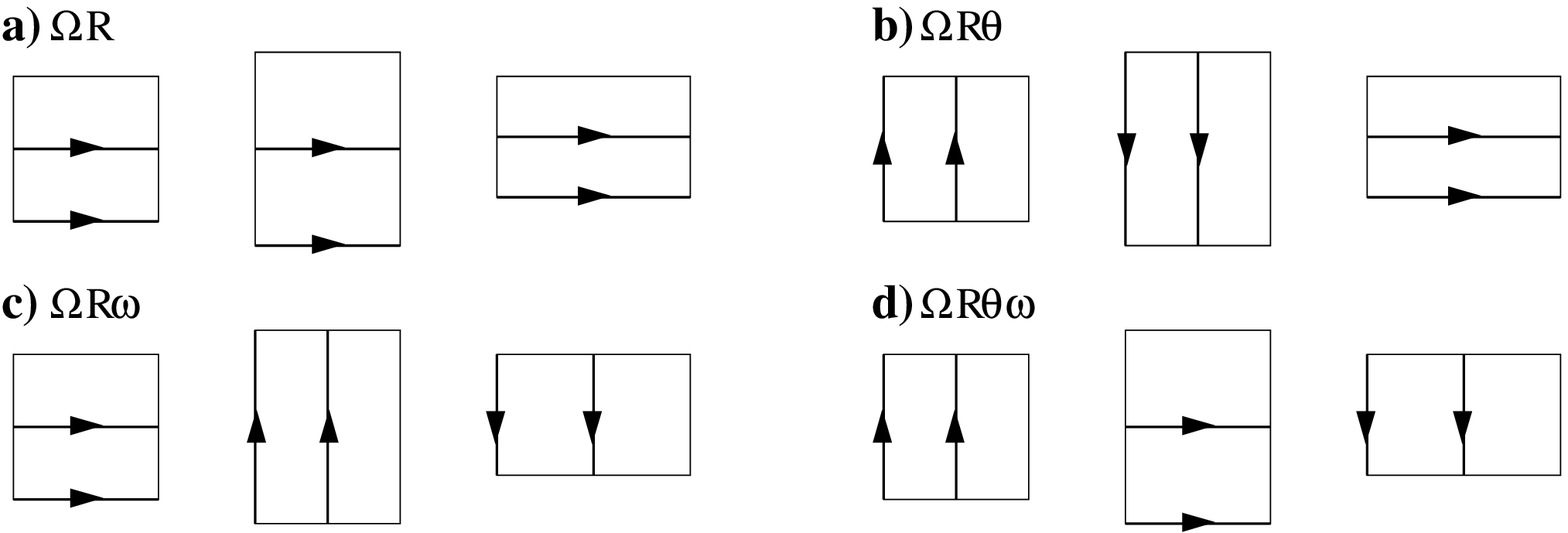}
\end{center}
\caption[]{\small O6-planes in the orientifold of $\IT^6/(\IZ_2\times
\IZ_2)$.}
\label{orient}
\end{figure}

We implement an orientifold projection by $\Omega R$, where
$\Omega$ is world-sheet parity, and $R$ acts as \beqa R:
(z_1,z_2,z_3) \to ({\ov z}_1,{\ov z}_2,{\ov z}_3). \eeqa There are then  four kinds of
orientifold 6-planes (O6-planes), associated with the actions of
$\Omega R$, $\Omega R\theta$, $\Omega R \omega$, and $\Omega
R\theta\omega$, as shown in Figure~\ref{orient}.

The orientifold group acts on the Chan-Paton indices of
the branes by the following
actions \beqa \gamma_{\theta,a} & = & \diag(i \id_{N_a/2},-i
\id_{N_a/2}\, ;
-i \id_{N_a/2},i \id_{N_a/2}) \nonumber \\
\nonumber \\
\gamma_{\omega,a} & = & \diag \left[ \pmatrix{0 & \id_{N_a/2} \cr
-\id_{N_a/2} & 0 } \; ; \;
\pmatrix{0 & \id_{N_a/2} \cr -\id_{N_a/2} & 0 } \right] \nonumber \\
\nonumber \\
\gamma_{\Omega R,a} & = & \pmatrix{ & & \id_{N_a/2} & 0 \cr & & 0
& \id_{N_a/2} \cr \id_{N_a/2} & 0 & & \cr 0 & \id_{N_a/2} & & \cr
} \eeqa The actions for the orbifold group form a projective
representation as explained in \cite{CSU2}.

\begin{figure}
\begin{center}
\centering
\epsfysize=7.5cm
\leavevmode
\epsfbox{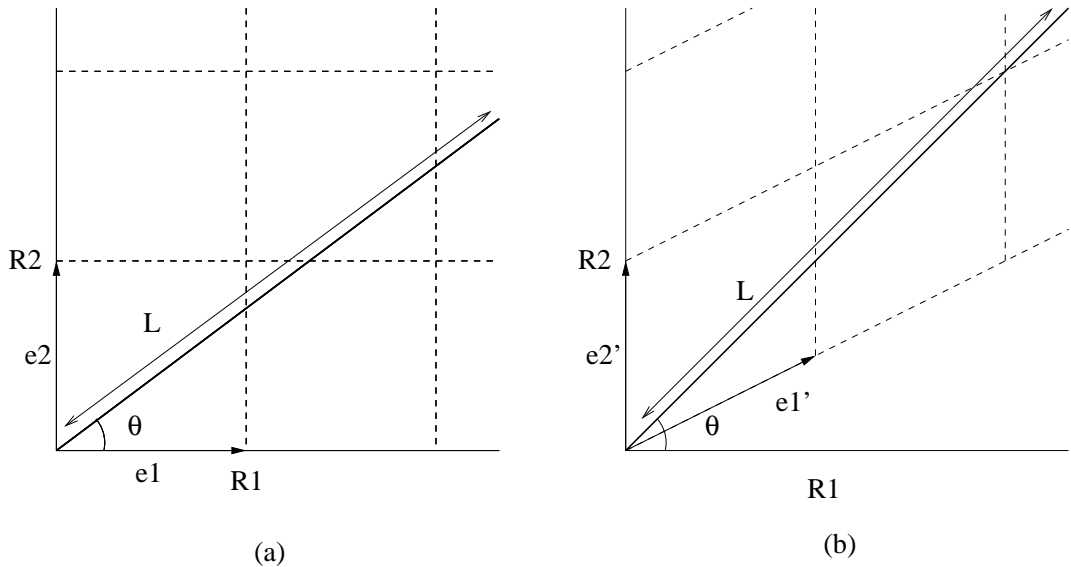}
\end{center}
\caption{\small The D6-branes wrap one-cycles on each two-torus.
The one-cycles make an angle $\theta$ with the $\Omega R$
orientifold plane which lies along the $R_1$-axis on all three
two-tori. The tori can be rectangular (a) or tilted (b).}
\label{torus}
\end{figure}

\begin{figure}
\begin{center}
\centering
\epsfysize=5.5cm
\leavevmode
\epsfbox{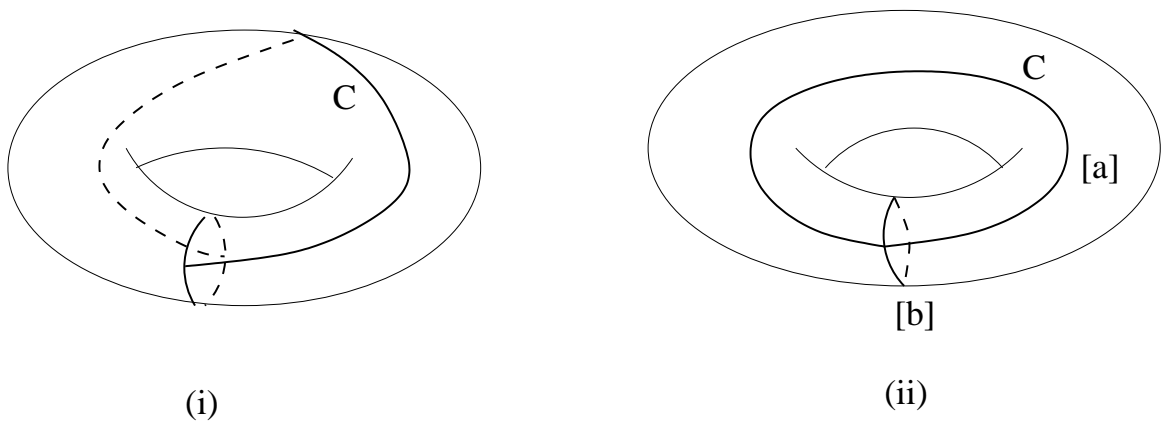}
\end{center}
\caption{\small The cycle $[a]+\frac{1}{2}[b]$, depicted as cycle C, is not closed for an
untilted torus (i). However, for a tilted torus (ii) the complex structure makes C a closed cycle $[a']$. }
\label{cycles}
\end{figure}

To cancel the RR charge of the O6-planes, we introduce D6-branes
wrapped on three-cycles that are products of one-cycles in each of
the three two-tori  (Figure \ref{torus}). Let $[a_i]$, $[b_i]$,
$i=1, 2, 3$, be a canonical basis of homology cycles. We consider
$K$ stacks of $N_a$ D6-branes, $a=1,\ldots, K$, wrapped on the
$n_a^i[a_i]+m_a^i[b_i]$ cycle in the $i^{th}$ two-torus. The
complex structure of the tori is arbitrary but it should be
consistent with the orientifold projection. The only allowed
possibilities then are shown in Figure \ref{torus}. We can either
have a rectangular torus (a) or a tilted torus (b) for which the
lattice vectors are $e_1'=e_1+e_2/2$, $e_2'=e_2$. If the above
homology basis refers to a rectangular torus, the cycle
$[a]+\frac{1}{2}[b]$ is not closed for a rectangular torus (Figure
\ref{cycles} (i)). However, it becomes closed for a tilted torus
because the complex structure compensates for the offset (Figure
\ref{cycles} (ii)). Therefore, a generic one-cycle on a
rectangular torus takes the form $n_a^i[a_i]+m_a^i[b_i]$, where
$n_a^i,\;\; m_a^i$ are integers, while on a tilted torus it takes
the form $n_a^i[a_i']+m_a^i[b_i]$, where $n_a^i,\;\; m_a^i$ are
again integers but $[a_i']=[a_i]+\frac{1}{2}[b_i]$ in terms of the
rectangular torus cycles. So the one-cycles on tilted tori can be
written as $n_a^i[a_i]+\tilde{m}_a^i[b_i]$, where
$\tilde{m}_a^i=m_a^i+n_a^i/2$ is a half integer. It is convenient
to describe rectangular and tilted tori cycles in a common
notation and to this end we introduce \beq l_{a}^{i}\equiv
m_{a}^{i},\;{\rm rectangular, }\;\;\;\;\;\; l_{a}^{i}\equiv
2\tilde{m}_{a}^{i}=2m_{a}^{i}+n_{a}^{i},\; {\rm tilted}. \eeq

With this definition we can label a generic one-cycle on either a rectangular or a tilted torus by
$(n_a^i,l_a^i)$. Note that for a tilted torus $l-n$ is necessarily even. In
addition, to avoid multiply wrapped branes we require that $m$
and $n$ are relatively coprime, for both tilted and untilted
tori.
Under an $\Omega R$ reflection a cycle $(n_a^i,l_a^i)$ is mapped to
$(n_a^i,-l_a^i)$. So, in order to implement the orientifold projection at the level of the
spectrum, for a stack of $N_a$ D6-branes along cycle $(n_a^i,l_a^i)$ we also need to include their
images with wrapping numbers $(n_a^i,-l_a^i)$.
For branes on top of the O6-planes we also count branes and their
images independently.

As discussed above, the homology three-cycles for stack $a$
of $N_a$ D6-branes and its
orientifold image $a'$ are given by

\beq
[\Pi_a]=\prod_{i=1}^{3}\left(n_{a}^{i}[a_i]+2^{-\beta_i}l_{a}^{i}[b_i]\right),\;\;\;
\left[\Pi_{a'}\right]=\prod_{i=1}^{3}\left(n_{a}^{i}[a_i]-2^{-\beta_i}l_{a}^{i}[b_i]\right)
\eeq where $\beta_i=0$ if the $i$th torus
is not tilted and $\beta_i=1$ if it is
tilted.  The homology three-cycles wrapped by the four orientifold
planes are
\beq
\begin{array}{clcl}
\Omega R: & [\Pi_1]=8[a_1]\times[a_2]\times[a_3],& \Omega R\omega: &
[\Pi_2]=-2^{3-\beta_2-\beta_3}[a_1]\times[b_2]\times[b_3]\\
\\ \Omega R\theta\omega: &
[\Pi_3]=-2^{3-\beta_1-\beta_3}[b_1]\times[a_2]\times[b_3], &
\Omega R\theta: &
[\Pi_4]=-2^{3-\beta_1-\beta_2}[b_1]\times[b_2]\times[a_3]
\end{array}\label{orienticycles}\eeq
and we define $[\Pi_{O6}]=[\Pi_1]+[\Pi_2]+[\Pi_3]+[\Pi_4]$.
The intersection numbers of the various
homology cycles  are easily computed using the
fact that the canonical homology one-cycles obey the Grassmann
algebra $[a_i][b_j]=-[b_j][a_i]=\delta_{ij},\;\;
[a_i][a_j]=[b_i][b_j]=0$. One finds \beq
\begin{array}{ll}
I_{ab}=[\Pi_a][\Pi_b]=2^{-k}\prod_{i=1}^3(n_a^il_b^i-n_b^il_a^i),&
I_{ab'}=[\Pi_a]\left[\Pi_{b'}\right]=-2^{-k}\prod_{i=1}^3(n_{a}^il_b^i+n_b^il_a^i)
\\\\
I_{aa'}=[\Pi_a]\left[\Pi_{a'}\right]=-2^{3-k}\prod_{i=1}^3(n_a^il_a^i),&
\\\\
\multicolumn{2}{l}{I_{aO6}=[\Pi_a][\Pi_{O6}]=2^{3-k}(-l_a^1l_a^2l_a^3+l_a^1n_a^2n_a^3+n_a^1l_a^2n_a^3+n_a^1n_a^2l_a^3)}
\end{array}
\label{intersections}\eeq where $k=\beta_1+\beta_2+\beta_3$ is the
total number of tilted tori.

\begin{figure}
\begin{center}
\centering
\epsfysize=5.5cm
\leavevmode
\epsfbox{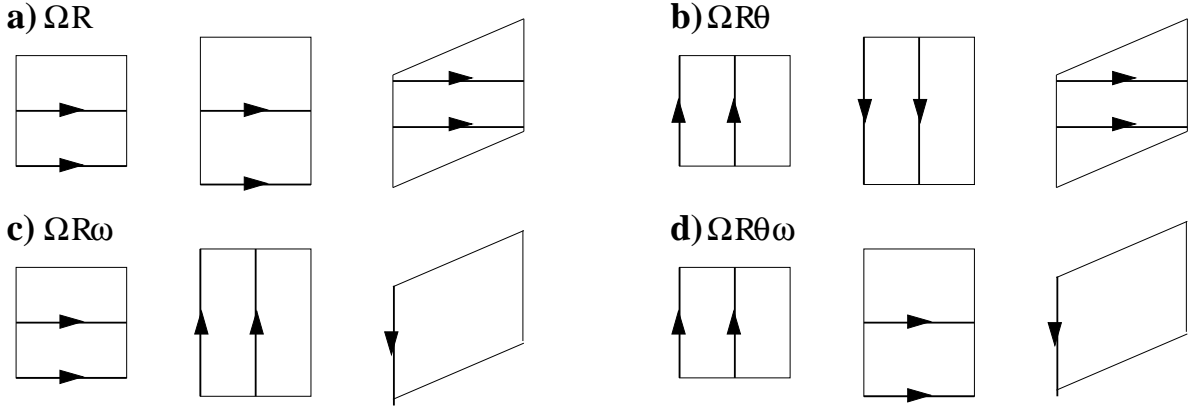}
\end{center}
\caption[]{\small O6-planes in the orientifold of $\IT^6/(\IZ_2\times
\IZ_2)$ where the third two-tori is tilted.}
\label{tiltorienti}
\end{figure}

As is shown in Figure \ref{orient} there are two orientifold
planes that wrap around each of the two non-contractible cycles in
a rectangular torus. So, if all tori are rectangular there are
eight orientifold planes of each type. For a tilted torus,
however, one of the two possible positions for the $[b]$-cycle is
lost (fig. \ref{tiltorienti}). So, depending on how many and which
tori are tilted, there could be less than eight orientifold planes
of the types $\Omega R\theta$, $\Omega R \omega$, and $\Omega
R\theta\omega$.
Equation
\ref{orienticycles} gives exactly the homology three-cycle of the four types of orientifold planes
{\em times} their multiplicity for an arbitrary number of tilted tori. Note that this normalization
is different from that used in \cite{CSU2}, where the multiplicity of the planes was not included in this definition of the cycles.
The new definition allows the tadpole cancellation and
supersymmetry conditions to be expressed in a form that is independent of the number of tilted tori. The results, of course,
are insensitive to which convention one uses.

The open string spectrum of these constructions for branes at
generic angles was discussed in detail in \cite{CSU2}. 
We summarize the results in table~\ref{matter}.\footnote{See Note Added 
at the end of the paper.}
Notice that because of the change of notation from \cite{CSU1,CSU2}, 
the multiplicities of $\Ysymm$ and $\Yasymm$ in the $aa' + a'a$ sector
have slightly different expressions.
\begin{table}
[htb] \footnotesize
\renewcommand{\arraystretch}{1.25}
\begin{center}
\begin{tabular}{|c|c|}
{\bf Sector} & {\bf Representation}
\phantom{more space inside this box} \\
\hline\hline
$aa$   & $U(N_a/2)$ vector multiplet  \phantom{more space inside this box}\\
       & 3 Adj. chiral multiplets  \phantom{more space inside this box} \\
\hline
$ab+ba$   & $I_{ab}$ $(\fund_a,\antifund_b)$ fermions  \phantom{more space inside this box} \\
\hline
$ab'+b'a$ & $I_{ab'}$ $(\fund_a,\fund_b)$ fermions \phantom{more space inside this box} \\
\hline $aa'+a'a$ & $-\frac 12 (I_{aa'} - \frac 12 I_{a,O6})\;\;
\Ysymm\;\;$ fermions  \phantom{more space inside this box} \\
          & $-\frac 12 (I_{aa'} + \frac 12 I_{a,O6}) \;\;
\Yasymm\;\;$ fermions \phantom{more space inside this box} \\
\end{tabular}
\end{center}
\caption{\small General spectrum on D6-branes at generic angles
(namely, not parallel to any O6-plane in all three tori). 
The spectrum is valid for both tilted and untilted tori.
The models may contain
additional non-chiral pieces in the $aa'$ sector and in $ab$,
$ab'$ sectors with zero intersection, if the relevant branes
overlap. In supersymmetric situations, scalars combine with the
fermions given above to form chiral supermultiplets.
\label{matter} }
\end{table}

\subsection{Tadpole conditions}

Cancellation of the Ramond-Ramond charge requires that the total
homology cycle, weighted by the D6-brane and O6-plane (-4 in
D6-brane units) charge, vanishes. That is \beq \sum_a N_a
[\Pi_a]+\sum_a N_a \left[\Pi_{a'}\right]-4[\Pi_{O6}]=0. \eeq It is
useful to introduce the products of wrapping numbers \beq
\begin{array}{rrrr}
A_a=-n_a^1n_a^2n_a^3 & B_a=n_a^1l_a^2l_a^3 & C_a=l_a^1n_a^2l_a^3 & D_a=l_a^1l_a^2n_a^3 \\
\tilde{A}_a=-l_a^1l_a^2l_a^3 & \tilde{B}_a=l_a^1n_a^2n_a^3 &
\tilde{C}_a=n_a^1l_a^2n_a^3 & \tilde{D}_a=n_a^1n_a^2l_a^3
\end{array}
\label{variables}\eeq It is then straightforward to rewrite this
as the set of equations \beq \sum_a N_a A_a=\sum_a N_a B_a=\sum_a
N_a C_a=\sum_a N_a D_a=-16. \eeq At this point we can introduce an
arbitrary number of branes wrapping cycles along the orientifold
planes, so called `filler branes', which contribute to the tadpole
conditions but trivially satisfy the supersymmetry conditions as
we shall see below. Table \ref{orientifold} shows the wrapping
numbers of the four O6-planes. Each of the O6-planes has only one
of $A,\,B,\,C,\,D$ equal to $-2^k$ and the rest are zero. So, if we
consider $N^{(1)}$ branes wrapped along the first orientifold
plane, $N^{(2)}$ along the second and so on, the tadpole
conditions are modified to
\begin{eqnarray}
 -2^k N^{(1)}+\sum_a N_a A_a=-2^k N^{(2)}+\sum_a N_a
B_a= \nonumber\\ -2^k N^{(3)}+\sum_a N_a C_a=-2^k N^{(4)}+\sum_a
N_a D_a=-16.
\end{eqnarray}Note that the tadpole conditions are symmetric in $A,\;B,\;C$ and
$D$.

\begin{table}
\footnotesize
\renewcommand{\arraystretch}{1.25}
\begin{center}
\begin{tabular}{|c|c|}
    & $(n^1,l^1)\times (n^2,l^2)\times (n^3,l^3)$ \phantom{some space inside this box}\\
\hline
    $\Omega R$& $(2^{\beta_1},0)\times (2^{\beta_2},0)\times (2^{\beta_3},0)$\phantom{some space inside this box} \\
    $\Omega R\omega$& $(2^{\beta_1},0)\times (0,2^{\beta_2})\times (0,-2^{\beta_3})$ \phantom{some space inside this box} \\
    $\Omega R\theta\omega$& $(0,2^{\beta_1})\times (2^{\beta_2},0)\times (0,-2^{\beta_3})$ \phantom{some space inside this box} \\
    $\Omega R\theta$& $(0,2^{\beta_1})\times (0,-2^{\beta_2})\times
    (2^{\beta_1},0)$ \phantom{some space inside this box} \\
\end{tabular}
\end{center}
\caption{Wrapping numbers of the four O6-planes.}
\label{orientifold}
\end{table}

\subsection{Conditions for supersymmetric brane configuration}
The condition for preserving ${\cal N}=1$, D=4 supersymmetry is
that the angle of rotation of any D-brane with respect to the orientifold
plane is an element of $SU(3)$, i.e., the matrix of rotation acting on
the complexified compact coordinates has unit determinant.
In
other words we require that $\theta_1+\theta_2+\theta_3=0 \;
\rm{mod}~2\pi$ where $\theta_i$ is the angle with the R-invariant
axis of the $i$th torus as shown in Figure \ref{torus}. This is
equivalent to $\sin(\theta_1+\theta_2+\theta_3)=0$, and
$\cos(\theta_1+\theta_2+\theta_3)>0$. We can easily express the
angles $\theta_i$ in terms of the one-cycle wrapping numbers on
the $i$th torus as \beq \sin\theta_i=\frac{2^{-\beta_i}l^i
R_2^i}{L^i(n^i,l^i)},\;\;\;\;\cos\theta_i=\frac{n^i
R_1^i}{L^i(n^i,l^i)},\eeq where
$L^i(n^i,l^i)=\sqrt{(2^{-\beta_i}l^i R_2^i)^2+(n^i R_1^i)^2}$ is
the length of the one-cycle wrapping the $i$th torus. Now one can
formulate the supersymmetry conditions in terms of the variables
introduced in equation (\ref{variables}). We obtain

\begin{eqnarray}
x_A\tilde{A}_a+x_B\tilde{B}_a+x_C\tilde{C}_a+x_D\tilde{D}_a=0
\nonumber\\\nonumber \\ A_a/x_A+B_a/x_B+C_a/x_C+D_a/x_D<0
\label{susyconditions}
\end{eqnarray} where $x_A=\lambda,\;
x_B=\lambda 2^{\beta_2+\beta3}/\chi_2\chi_3,\; x_C=\lambda
2^{\beta_1+\beta3}/\chi_1\chi_3,\; x_D=\lambda
2^{\beta_1+\beta2}/\chi_1\chi_2$ and $\chi_i=(R_2/R_1)_i$ are the
complex structure moduli. The positive parameter $\lambda$ has
been introduced to put all the variables $A,\,B,\,C,\,D$ on an equal
footing. However, among the $x_i$ only three of them are
independent and in the end of the calculation we should express
three of them in terms of the fourth. In contrast to the tadpole
conditions, supersymmetry constrains each stack of branes
individually and can therefore be used to classify all possible
brane configurations that preserve supersymmetry. The problem of
model building is then enormously simplified since there is only a
finite number of building blocks one can possibly combine to
construct consistent models.

We have seen that neither the tadpole nor the supersymmetry
conditions differentiate among $A,\;B,\;C$ or $D$. Equivalence of
$B,\;C$ and $D$ simply follows from the equivalence of the three
tori. The equivalence between $A$ and the rest three variables is
related to the configuration of the orientifold planes. One can
interchange $A$ with one of the other variables by exchanging $l$
with $n$ for one torus and simultaneously replacing $l$ with$-n$
and $n$ with $-l$ for a second torus. Indeed this is precisely the
transformation that exchanges the $\Omega R$-plane with one of the
other three orientifold planes. Moreover, it is the presence of
the orientifold planes that breaks the symmetry between the two
axes of the two-tori and hence differentiates $A,\;B,\;C$ and $D$
from  $\tilde{A},\;\tilde{B},\;\tilde{C}$ and $\tilde{D}$. It is
therefore clear that the symmetries between these variables will
be different for different orbifold groups and can generically be
determined from the configuration of orientifold planes.

\section{Classification of supersymmetric brane configurations}

Since at most one of $l$ or $n$ can be zero for each torus any
brane configuration belongs to one of four possible classes. It
can have a zero wrapping number in all three, two, one or none of
the tori. We examine each case separately.

\begin{description}
\item[I] {\bf Three zeros}\\
        Supersymmetry requires that at least one of $A,\;B,\;C$ or $D$ be non-zero, in fact negative.
        So either all $l$'s are zero or two $n$'s and an $l$ are
        zero. The value of the non-zero wrapping numbers is $\pm
        1$ for a non-tilted torus and $\pm 2$ for a tilted torus as
        required to avoid multiply wrapped cycles.
        Their relative sign is such that the non-vanishing
        product, $A,\;B,\;C$ or $D$, is negative. We then find
        exactly four possible brane configurations, namely one of
        the orientifold planes and its three images under the
        other three orientifold planes. Therefore, up to
        $O6$-plane reflections there are just four inequivalent
        brane configurations, each one parallel to one of the
        orientifold planes. These are precisely the `filler'
        branes which, as advertised, do not constrain the moduli
        and we have already included the
        effect of an arbitrary number of each of them in the
        tadpole conditions.

\item[II] {\bf Two zeros}\\
        Each wrapping number enters in two of the products $A,\;B,\;C$ and
        $D$ and two of $\tilde{A},\;\tilde{B},\;\tilde{C}$ and
        $\tilde{D}$. Since not both $l$ and $n$ can be zero for a
        given torus we necessarily have one and only one of $A,\;B,\;C$
        or $D$ and one of $\tilde{A},\;\tilde{B},\;\tilde{C}$ or $\tilde{D}$
        non-vanishing. But then, supersymmetry requires that the
        latter vanishes and hence there is a third zero which
        contradicts the hypothesis of two zero wrapping numbers.
        Therefore there are no supersymmetric configurations with
        two zero wrapping numbers.
\item[III] {\bf One zero}\\
        Here precisely two of $A,\;B,\;C$ and
        $D$ and two of $\tilde{A},\;\tilde{B},\;\tilde{C}$ and
        $\tilde{D}$ are zero. There are six cases depending on
        which wrapping number we choose to be zero. As an example
        let $n^1=0$ so that $A=B=\tilde{C}=\tilde{D}=0$. The
        identity $CD=-\tilde{A}\tilde{B}$ together with the
        supersymmetry conditions then imply that \beq C<0,\;D<0,\;
        \tilde{A}\tilde{B}<0,\;\; x_A/x_B=-\tilde{B}/\tilde{A}.
        \eeq We refer to this type of supersymmetric brane
        configuration as `type III'.
\item[IV]{\bf No zeros}\\
        Since no wrapping number is zero here we have
        $A\tilde{A}=B\tilde{B}=C\tilde{C}=D\tilde{D}={\rm
        constant}\neq 0$. The supersymmetry conditions then require
        that one and only one of $A,\;B,\;C$ or $D$ is positive and
        \beq x_A/A+x_B/B+x_C/C+x_D/D=0. \eeq We refer to this type
        of supersymmetric configuration as `type IV'.
\end{description}

We are now in a position to start looking for consistent models by
combining stacks of type III and type IV branes such that the
tadpole conditions, as well as the supersymmetry conditions for
the moduli are satisfied. Before proceeding however it is
instructive to think about the compatibility of different type III
and type IV configurations. Consider first two stacks of type IV,
one with, say, $A_1>0$ and one with $B_2>0$. Each of them gives an
equation for the moduli which can only have common solution
provided \beq A_1B_2<B_1A_2. \eeq The case $A_1>0,\;A_2>0$
requires that two of $|B_1/A_1|,\;|C_1/A_1|,\;|D_1/A_1|$ are
smaller than their counterparts for the second stack and the third
is bigger, or vice versa. Next consider one stack of type IV
with $A_1>0$ and a second one of type III. These two configurations
are always compatible unless the type III configuration has
$A_2=B_2=0$ in which case $x_A/x_B=-\tilde{B}_2/\tilde{A}_2$ and a
common solution is only possible if  \beq
|A_1/B_1|<-\tilde{B}_2/\tilde{A}_2.\eeq Finally consider two
stacks of type III. These are generically compatible unless they
fix the same moduli ratio, say $x_A/x_B$, to different values.

We could go on and consider general compatibility conditions among
three stacks of branes and so on but the number of cases to be
considered increases considerably and it seems more economical to
examine case by case as it arises. However it is important to
realize at this point that generically three stacks of type III
and/or type IV configurations completely fix the three moduli and
so there is no freedom of adding a fourth stack of branes. We
therefore consider only configurations of up to three stacks of
type III and/or type IV branes. Nevertheless, one could imagine the
possibility of a consistent model
with more than three stacks of branes not parallel to the
orientifold planes and with subsets of these stacks giving the
same equations for the moduli. A special case of this possibility
is realized when a single stack of branes is split into two
parallel stacks by moving the position of the three-cycle in one
or more of the tori. This special case is taken into account by
simply treating the number of branes in each stack as an arbitrary
parameter to be fixed by the tadpole conditions as we do in the
analysis below. However, since all images of a given brane
configuration under the $O6$-planes contribute in exactly the same
way in the tadpole and supersymmetry conditions (any $O6$-plane
reflection leaves $A,\;B,\;C$ and $D$ invariant and changes the
sign of $\tilde{A},\;\tilde{B},\;\tilde{C}$ and $\tilde{D}$), our
analysis automatically accounts for this slightly more general
case as well, which cannot be obtained by the splitting mechanism
described above. To completely exclude the possibility of more
than three stacks not parallel to the orientifold planes one needs
to show that a fourth stack necessarily gives a moduli equation that cannot be
written as a linear combination of those of the first three stacks.

\section{Search for supersymmetric three-family $SU(5)$ GUTS}

In the standard SU(5) GUT, the quarks and leptons are embedded in
the {\bf 10}
and $\overline{\bf 5}$ of SU(5).
We are therefore interested in models
containing a stack with at least five branes which has three
copies of antisymmetric matter, i.e. $n_{\Yasymm}=\pm 3$. Let us
investigate the consequences of this constraint. From Table
\ref{matter} and Equation (\ref{intersections}) one sees that \beq
n_{\Yasymm}=2^{1-k}[(2A-1)\tilde{A}-\tilde{B}-\tilde{C}-\tilde{D}],\;\;
n_{\Ysymm}=2^{1-k}[(2A+1)\tilde{A}+\tilde{B}+\tilde{C}+\tilde{D}]\eeq
These are symmetric in $A,\;B,\;C$ and $D$ due to the identity
$A\tilde{A}=B\tilde{B}=C\tilde{C}=D\tilde{D}$. For a filler brane
both these expressions vanish identically and so the U(5) stack of
branes must be either type III or type IV. We examine the latter
case first.

\subsection{Type IV brane} Without loss of generality we take
$A>0,\;\tilde{A}>0$. Then,\beq
n_{\Yasymm}=2^{1-k}[(2|A|-1)|\tilde{A}|+|\tilde{B}|+|\tilde{C}|+|\tilde{D}|]\geq
2^{3-k}.\eeq We immediately conclude that $k=2$ or $k=3$.

\begin{description}
\item[k=2] Here
$(2|A|-1)|\tilde{A}|+|\tilde{B}|+|\tilde{C}|+|\tilde{D}|=6$ and we
need to consider the four-partitions of 6. There are just two,
6=3+1+1+1=2+2+1+1. We identify the following inequivalent
possibilities

\begin{description}
\item[(i)] $(2|A|-1)|\tilde{A}|=3,\;|\tilde{B}|=|\tilde{C}|=|\tilde{D}|=1$
\item[(ii)]$(2|A|-1)|\tilde{A}|=1,\;|\tilde{B}|=3,\;|\tilde{C}|=|\tilde{D}|=1$
\item[(iii)]$(2|A|-1)|\tilde{A}|=|\tilde{D}|=1,\; |\tilde{B}|=|\tilde{C}|=2$
\item[(iv)]$(2|A|-1)|\tilde{A}|=|\tilde{B}|=2,\;|\tilde{C}|=|\tilde{D}|=1$
\end{description}
The first three possibilities are trivially excluded since any
non-trivial factor must appear simultaneously in two of the
products $A,\;B,\;C$ and $D$ and in two of
$\tilde{A},\;\tilde{B},\;\tilde{C}$ and $\tilde{D}$. However, the
last possibility passes this elementary test with the solution
$(n^1,l^1)\times (n^2,l^2)\times (n^3,l^3)=(-1,-2)\times
(-1,-1)\times (-1,-1)$ or $(A,B,C,D)=(1,-1,-2,-2)$ and
$(\tilde{A},\tilde{B},\tilde{C},\tilde{D})=(2,-2,-1,-1)$. Since
for a tilted torus $n-l$ is even we see that the second and third
tori are tilted while the first is not. Next we consider the
tadpole conditions. To obtain a U(5) gauge group we need a stack
of at least 10 coincident branes but we could have more and then
separate the additional branes by the splitting mechanism
discussed above. Allowing for $N$ extra branes we obtain
\begin{eqnarray}
 \sum_a N_a A_a=2(-13-N/2-2 N^{(1)}),\;\;\sum_a N_a B_a=2(-3+N/2+2 N^{(2)}), \nonumber\\
 \sum_a N_a C_a=2(2+N+2N^{(3)})>0,\;\;\sum_a N_a D_a=2(2+N+2N^{(4)})>0.
\end{eqnarray}
The only way to satisfy the last two conditions is to add at least
two extra stacks of type IV, one with $C_1>0$ and a second with
$D_2>0$. As we have seen above this can only be consistent
provided $C_1D_2<D_1C_2$. We have now increased the total number
of non-trivial stacks to three and so we assume we cannot obtain a
solution by adding more stacks. The last two tadpole conditions
then imply $C_1D_2>D_1C_2$ which contradicts the above inequality.
Therefore there is no solution for k=2.

\item[k=3]Here
$(2|A|-1)|\tilde{A}|+|\tilde{B}|+|\tilde{C}|+|\tilde{D}|=12$ and
we need to consider the four-partitions of 12. There are fifteen
such partitions but since all tori are tilted only partitions with
all terms either even or odd must be kept. There are just five of
those, 12=7+3+1+1=4+4+2+2= 3+3+3+3=5+5+1+1=5+3+3+1. The first
three can easily be excluded since they cannot be realized by the
coefficients $\tilde{A},\;\tilde{B},\;\tilde{C}$ and $\tilde{D}$.
Each of the last two partitions gives a unique solution as follows
\begin{description}
\item[(i)]$(A,B,C,D)=(1,-1,-5,-5),\;\;(\tilde{A},\tilde{B},\tilde{C},\tilde{D})=(5,-5,-1,-1)$
\item[(ii)]$(A,B,C,D)=(3,-3,-1,-1),\;\;(\tilde{A},\tilde{B},\tilde{C},\tilde{D})=(1,-1,-3,-3)$
\end{description} The first case can be shown to be impossible by
an argument similar to that used to exclude case (iv) for k=2
above. The key property is that the tadpole conditions again
require the addition of two stacks of type IV which cannot satisfy
simultaneously the tadpole and supersymmetry conditions.

The second case however is much more interesting. Remarkably
enough it is the only possible configuration that has
simultaneously 3 antisymmetric ({\bf 10}) and no symmetric ({\bf
15}) multiplets and therefore comes closer to a realistic SU(5)
GUT. Unfortunately it turns out, after considerably more effort
than in the previous cases, that again there is no solution, at
least with up to three non-trivial stacks of branes. The tadpole
conditions in this case are
\begin{eqnarray}
 \sum_a N_a A_a=-2(23+3N/2-4 N^{(1)}),\;\;\sum_a N_a B_a=2(7+3N/2+4N^{(2)})>0, \nonumber\\
 \sum_a N_a C_a=-2(3-N/2-4N^{(3)}),\;\;\sum_a N_a
 D_a=-2(3-N/2-4N^{(4)}).\hspace{0.6cm}
\end{eqnarray}The second condition requires the addition of a
second stack of type IV with $B_1>0$. Compatibility with the U(5)
stack then implies $|B_1|<|A_1|$. Moreover, the last two tadpole
conditions require $N^{(3)}=N^{(4)}=0$ and $N=0,\;2,\;4$.
Inserting then the first two tadpole conditions into the above
inequality one obtains $4-N^{(1)}-N^{(2)}>0$. That is either both
$N^{(1)}$ and $N^{(2)}$ are zero or one of them is zero and the
other 2. Considering then the cases $N=0,\;2,\;4$ separately one
easily sees that there is no solution with just two non-trivial
stacks unless we are willing to accept the limiting case
$|B_1|=|A_1|$ corresponding to setting two of the moduli to zero.
A similar analysis can then be repeated for three non-trivial
stacks. There are quite a few cases one needs to consider since
the third brane is not constrained a priori and can be either type
III or type IV. In either case the supersymmetry conditions from
all three stacks become very constraining and one can
systematically show that there is no solution, except for limiting
cases similar to the one mentioned above. We have therefore shown
that there is no three-family U(5) model from type IV
configurations. We turn next to the alternative case of type III
configurations.
\end{description}

\subsection{Type III brane}

Without loss of generality we take $A=B=\tilde{C}=\tilde{D}=0$
i.e. $n^1=0$. Then \beq n_{\Yasymm}=-n_{\Ysymm}=\pm
2^{1-k}(|\tilde{A}|-|\tilde{B}|)\eeq and so we need
$|\tilde{A}|-|\tilde{B}|=\pm 3\times 2^{k-1}$. Hence $k\geq1$. Note that
for type III brane we necessarily have the same number of symmetric and antisymmetric multiplets which means
that any model we construct in this categoty will have three copies of {\bf 15}-plets. The
tadpole conditions are
\begin{eqnarray}
 -2^k N^{(1)}+\sum_a N_a A_a=0,\hspace{2.6cm}-2^k N^{(2)}+\sum_a N_a B_a=0,\hspace{3.0cm} \nonumber\\
 -2^k N^{(3)}+\sum_a N_a C_a=(10+N)|C|-16,\;\;-2^k N^{(4)}+\sum_a N_a
 D_a=(10+N)|D|-16.\hspace{0.5cm}
\end{eqnarray} If $|C|=|D|=1$ then $|\tilde{A}|=|\tilde{B}|=1$ as
well which violates the three families condition. Therefore at
least one of $|C|$ and $|D|$ is greater than zero. There are two
cases
\begin{description}
\item[(i)]$|C|>1$ and $|D|>1$
\item[(ii)]$|C|=1$ and $|D|>1$
\end{description}
For the first case the tadpole conditions force us to add two
stacks of type IV branes, one with $C_1>0$ and one with $D_2>0$. One
easily concludes then that this case is impossible following an
argument similar to the ones used to exclude case (iv), k=2 and
case (i), k=3 in the previous section. For the second case we
first note that $|C|=1$ implies $|l^1|=1$ and since $n^1=0$ we
conclude that the first torus cannot be tilted. Hence, either k=1
or k=2. The tadpole conditions require that we add a second stack
of type IV branes with $D_1>0$ and we may add a third stack which must
necessarily have $C_2\leq 0$. $C_2> 0$ is excluded for the same
reason  as case (i). In any case one finds
$|\tilde{A}|=|l^2|,\;|\tilde{B}|=|n^3|,\;|D|=|l^2n^3|$. The three
families condition then implies $|D|=m(m+3\times 2^{k-1})\geq 4$,
where $m$ is a positive integer. At this point one can start a
systematic search for solutions by first looking at the case of
two stacks of non-trivial branes and then considering all
possibilities for a third stack. We have performed such a
systematic search and we list all the models found, together with their spectra,
in the appendix. We
have found  all 10 possible solutions for the case of two stacks
of non-trivial branes (models I) and all 149 solutions with a
third stack of type III (modes II, III and IV). We also list one
solution with a third stack of type IV (model V). It is in general
much harder to systematically look for solutions of this class
mainly because there are more variables to be fixed and the
supersymmetry conditions, although more constraining, are harder
to implement as a useful criterion to be used in the search and
must instead be checked in the end. Finally, we would like to
mention that models I differ qualitatively from the rest in that
they have one modulus free. All other models completely fix all
three moduli.

\section{Conclusions}

In this paper, we have explored the possibility of constructing realistic
$SU(5)$ Grand Unified string models from Type IIA orientifolds on $T^6/{\bf Z}_2 \times {\bf Z}_2$.
Due to the strong constraints from supersymmetry and tadpole cancellations,
we found that within this construction there are
no three-family supersymmetric $SU(5)$ models that
are free of  {\bf 15}-plets.
We then relax our criteria by allowing the appearance of {\bf 15}-plets
and systemically study the three-family supersymmetric $SU(5)$ models constructed in this framework.
The models are not fully realistic as there are a number of phenomenological
challenges that they have to face.
First of all,
under $SU(3) \times SU(2) \times U(1)_Y$, the {\bf 15} representation
decomposes
 as follows:
\begin{eqnarray}
SU(5) &\supset& SU(3) \times SU(2) \times U(1)_Y \nonumber \\
{\bf 15} &=& ({\bf 6},{\bf 1})(-\frac{2}{3})
+ ({\bf 1},{\bf 3})(+1) + ({\bf 3},{\bf 2})(+\frac{1}{6})
\end{eqnarray}
Therefore, the models we consider here contain in addition to
the Standard Model particles some exotic chiral matter fields.
Furthermore, in the brane construction, the Standard Model particles
are not only charged under
the $SU(5)$ gauge group but also an additional $U(1)$ which is the
center of mass motion of the stack of five D-branes (so the gauge group
is actually $U(5)$).
In the minimal $SU(5)$ model\footnote{In models with Higgs fields
transforming in
higher dimensional reps. of SU(5)
such as {\bf 45} rep., the fermion masses can come from
operators other than {\bf 10}~{\bf 10}~{\bf 5}$_H$ and
{\bf 5}~{\bf 10}~$\overline{\bf 5}_H$. However,
the perturbative open string sector does not give rise to
such higher dimensional
representations.}
 in which the only Higgs fields are
in the {\bf 24} and {\bf 5} representations,
the fermion masses come from the Yukawa couplings:
{\bf 10}~{\bf 10}~{\bf 5}$_H$ and
$\overline{\bf 5}$~{\bf 10} $\overline{\bf 5}_H$ where the subscript $H$
stands for the Higgs fields. While the operator $\overline{\bf 5}$~{\bf 10} $\overline{\bf 5}_H$ is allowed by
$U(1)$ charge conservation, the $U(1)$ charge carried by
{\bf 10}~{\bf 10}~{\bf 5}$_H$
is non-zero and hence 
forbidden\footnote{The {\bf 10}~{\bf 10}~{\bf 5}$_H$
coupling vanishes for the same reason in the non-supersymmetric $SU(5)$
model in \cite{bklo}.}.

One of the motivations for constructing GUT models in the
framework of intersecting D-branes is to explore if there are some
novel ways of solving some of the long-standing problems in GUTs,
e.g., the doublet-triplet splitting problem \cite{RandallCsaki}.
The orientifold models considered here when lifted to M theory
correspond to $G_2$ compactifications. Therefore, it would be
interesting to see if the mechanism suggested in
\cite{G2-23-split} can be applied to these models. The basic idea
in \cite{G2-23-split} is that the GUT symmetry can be broken by
Wilson lines in such a way that the doublet and the triplet have
different discrete quantum numbers. The discrete symmetry forbid a
mass term for the doublet whereas the mass term for the triplet is
allowed. Therefore, the doublet could remain light even though the
triplet receives a GUT scale mass. However, the Wilson lines in
the present context are continuous rather than discrete. Only for
some special choice of the continuous parameters of the
Wilson lines do we obtain the aforementioned discrete quantum
numbers.

Although the models we presented are not fully realistic, the
techniques that we developed in analyzing the supersymmetric
constraints and tadpole cancellations could easily be applied to
the search for realistic models in other orientifold constructions
(e.g., $T^6/\Gamma$ where $\Gamma$ is a discrete symmetry of
$SU(3)$ other than ${\bf Z}_2 \times {\bf Z}_2$). The fact that we
treat the tilted and rectangular tori in a symmetric manner
greatly simplifies the search for solutions to the constraints.
For simplicity, in the search for Standard-like models in
\cite{CSU1,CSU2,CSU3}, we assume that the angles that the
D6-branes are rotated with respect to the orientifold plane take
the form of $(\theta_1,\theta_2,0)$, $(0,\theta_2,\theta_3)$ or
$(\theta_1,0,\theta_3)$. It would be interesting to explore other
realistic Standard-like models from more general D6-brane
configurations. We hope to return to these problems in the future.

\section*{Acknowledgments}

We would like to thank Michael Douglas,
Paul  Langacker, and  Angel Uranga for useful
discussions.
MC would like to thank the  New
Center for Theoretical Physics at Rutgers University  for hospitality and
support  during the course of this work.
GS thanks the Kavli Institute for Theoretical Physics for hospitality
during the final stage of this work.
Research supported in part by DOE grant DOE-FG02-95ER40893 (MC, IP), NATO
linkage  grant No. 97061 (MC), National Science Foundation
Grant  No. INT02-03585 (MC, IP, GS),
Class of 1965 Endowed Term
Chair (MC), and funds from the University of Wisconsin (GS).
\vspace{0.5in}

{\it Note Added}

After this article was published we noticed that the sign of the multiplicities
of the symmetric and antisymmetric representations in Table 1 should be reversed. 
That is there are $\frac 12 (I_{aa'} - \frac 12 I_{a,O6})\,\,
\Ysymm\;\;$ fermions  and  $\frac 12 (I_{aa'} + \frac 12 I_{a,O6}) \,\,
\Yasymm\;\;$ fermions in the $aa'+a'a$ sector. As a consequence,
 the sign of the multiplicities of the symmetric and antisymmetric representations
should be reversed for all models in the Appendix, but all models remain
valid solutions to the constraints we imposed.

\newpage

\begin{center}
\Large{\bf Appendix}
\end{center}

In the following,
we tabulate all the models we have found together with their spectra. $a, b, c$ denote the
stacks of D-branes not parallel to the orientifold planes, giving $U(N_a/2)$ gauge group, while $a', b', c'$ denote
their $\Omega R$ image. 1,2,3,4 denote filler branes respectively along the $\Omega R$, $\Omega R\omega$, $\Omega R\theta\omega$ and $\Omega R\theta$
orientifold plane, resulting in a $USp(N^{(i)})$ gauge group. $N$ is the number of branes in each stack. The
third column shows the wrapping number of the various branes. Although we
do not specify in each model how many and which tori are tilted,
this can be seen most easily from the wrapping numbers of the O6-planes. $(2,0)$ or $(0,2)$ signify a tilted torus
(see Table \ref{orientifold}). For the very few cases where there are no filler branes (e.g. model II.1.1) we remind the reader
that at least one torus is tilted for all these models and for a tilted torus $n$ and $l$ are either both odd or both even.
The intersection numbers between the various stacks are given in the remaining columns to the right. For example, the intersection
number $I_{ac}$ between stacks $a$ and $c$ is found in row $a$ column $c$. For convenience we also list the relation among the moduli
imposed by the supersymmetry conditions, as well as the gauge group for each model. The numbering of the models reflects the order in which they were found in a systematic search and, to the extent this was possible, how closely the various models are related to one another.
In particular, models I have two stacks of branes not parallel to the orientifold planes, models II, III and IV have a third stack
of type III branes and model V has a third stack of type IV branes. In some cases (e.g. for models III.4.1-56)
we have parametrized a number of closely related models (56 in this case) with one or more  non-negative integers
which must satisfy definite constraints, e.g. $l_c^1 l_c^2=24-N^{(4)}$ in this example.
These constraints are also shown in the tables.
\vspace{0.5in}
\scriptsize

\begin{tabular}{|c||c|c||c|c|c|c|c|}
\hline
    \rm{model} I.1 & \multicolumn{7}{c|}{$U(5)\times U(1)\times USp(6)$}\\
\hline\hline \rm{stack} & $N$ & $(n^1,l^1)\times (n^2,l^2)\times
(n^3,l^3)$ & $n_{\Ysymm}$  & $n_{\Yasymm}$ & $b$ & $b'$ & 1 \\
\hline
    $a$&  10& $(0,-1)\times (1,4)\times (1,1)$ & 3 & -3  & 0 & 0 & 4\\
    $b$&   2& $(-2,3)\times (-1,4)\times (1,1)$ & 27 & 69 &- &- & -12\\
\hline
    1&   6& $(1,0)\times (1,0)\times (2,0)$ & \multicolumn{5}{c|}{$x_B=4x_A, \; x_A+x_C/3=x_D/12$}\\
\hline
\end{tabular}

\begin{tabular}{|c||c|c||c|c|c|c|c|c|c|}
\hline
    \rm{model} I.2 & \multicolumn{9}{c|}{$U(5)\times U(1)\times USp(4)\times USp(2)\times USp(6)$}  \\
\hline\hline \rm{stack} & $N$ & $(n^1,l^1)\times (n^2,l^2)\times
(n^3,l^3)$ & $n_{\Ysymm}$ & $n_{\Yasymm}$  & $b$ & $b'$ & 1 & 2 & 4\\
\hline
    $a$&  10& $(0,-1)\times (1,1)\times (4,1)$ & -3 & 3  & -8  & -12 & 1 & -4 & 0 \\
    $b$&   2& $(-2,3)\times (-1,3)\times (2,1)$ & 47 & 97  &-  &- & -9& -6 & 2\\
\hline
    1&   4& $(1,0)\times (2,0)\times (1,0)$ & \multicolumn{7}{c|}{$x_A=4x_B, \; 7x_B+2x_C=x_D/3$}\\
    2&   2& $(1,0)\times (0,2)\times (0,-1)$& \multicolumn{7}{c|}{}\\
    4&   6& $(0,1)\times (0,-2)\times (1,0)$ & \multicolumn{7}{c|}{}\\
\hline
\end{tabular}

\begin{tabular}{|c||c|c||c|c|c|c|c|c|}
\hline
    \rm{model} I.3 & \multicolumn{8}{c|}{$U(5)\times U(1)\times USp(2)\times USp(2)$} \\
\hline\hline \rm{stack} & $N$ & $(n^1,l^1)\times (n^2,l^2)\times
(n^3,l^3)$ & $n_{\Ysymm}$ & $n_{\Yasymm}$ & $b$ & $b'$ & 1 & 2 \\
\hline
    $a$&  10& $(0,-1)\times (1,7)\times (1,1)$ & 3 & -3 & 10 & 8 & 7 & -1 \\
    $b$&   2& $(-2,3)\times (-1,3)\times (3,1)$ & 37 & 71  &-  &- & -9& -9 \\
\hline
    1&   2& $(1,0)\times (2,0)\times (2,0)$ & \multicolumn{6}{c|}{$x_B=7x_A, \; 4x_A+x_C=x_D/9$}\\
    2&   2& $(1,0)\times (0,2)\times (0,-2)$ & \multicolumn{6}{c|}{}\\
\hline
\end{tabular}

\begin{tabular}{|c||c|c||c|c|c|c|c|c|}
\hline
    \rm{model} I.4 & \multicolumn{8}{c|}{$U(5)\times U(1)\times USp(2)\times USp(24)$}    \\
\hline\hline \rm{stack} & $N$ & $(n^1,l^1)\times (n^2,l^2)\times
(n^3,l^3)$ & $n_{\Ysymm}$ & $n_{\Yasymm}$ & $b$ & $b'$ & 1 & 4 \\
\hline
    $a$&  10& $(0,-1)\times (1,4)\times (1,1)$ & 3 & -3 & 16 & 0 & 4 & 0 \\
    $b$&   2& $(-2,3)\times (-1,4)\times (3,1)$ & 101 & 187&- &- & -12& 2\\
\hline
    1&   2& $(1,0)\times (1,0)\times (2,0)$ & \multicolumn{6}{c|}{$x_B=4x_A, \; 2x_A+x_C=x_D/18$}\\
    4&  24& $(0,1)\times (0,-1)\times (2,0)$ & \multicolumn{6}{c|}{}\\
\hline
\end{tabular}

\begin{tabular}{|c||c|c||c|c|c|c|c|c|}
\hline
    \rm{model} I.5 & \multicolumn{8}{c|}{$U(5)\times U(1)\times U(2)\times USp(2)\times USp(2)$}    \\
\hline\hline \rm{stack} & $N$ & $(n^1,l^1)\times (n^2,l^2)\times
(n^3,l^3)$ & $n_{\Ysymm}$ & $n_{\Yasymm}$ & $b$ & $b'$ & 1 & 2 \\
\hline
    $a$&  10+2& $(0,-1)\times (1,4)\times (1,1)$ & 3 & -3 & 7 & 2 & 4 & -1 \\
    $b$&   4& $(-1,1)\times (-1,3)\times (3,1)$ & 4 & 32 &- &- & -3& -3 \\
\hline
    1&   2& $(1,0)\times (1,0)\times (2,0)$ & \multicolumn{6}{c|}{$x_B=4x_A, \; 5x_A+3x_C=x_D/3$}\\
    2&   2& $(1,0)\times (0,1)\times (0,-2)$ & \multicolumn{6}{c|}{}\\
\hline
\end{tabular}

\begin{tabular}{|c||c|c||c|c|c|c|c|c|}
\hline
    \rm{model} I.6 & \multicolumn{8}{c|}{$U(5)\times U(1)\times U(2)\times USp(24)\times USp(8)$}    \\
\hline\hline \rm{stack} & $N$ & $(n^1,l^1)\times (n^2,l^2)\times
(n^3,l^3)$ & $n_{\Ysymm}$ & $n_{\Yasymm}$ & $b$ & $b'$ & 1 & 4 \\
\hline
    $a$&10+2& $(0,-1)\times (1,4)\times (1,1)$ & 3 & -3 & 8 & 0 & 4 & 0 \\
    $b$&   4& $(-1,1)\times (-1,4)\times (3,1)$ & 6 & 42 &- &- & -4& 1\\
\hline
    1&   2& $(1,0)\times (1,0)\times (2,0)$ & \multicolumn{6}{c|}{$x_B=4x_A, \; 4x_A+3x_C=x_D/4$}\\
    4&   8& $(0,1)\times (0,-1)\times (2,0)$& \multicolumn{6}{c|}{} \\
\hline
\end{tabular}

\begin{tabular}{|c||c|c||c|c|c|c|c|}
\hline
    \rm{model} I.7 & \multicolumn{7}{c|}{$U(5)\times U(1)\times U(1)\times USp(4)$}    \\
\hline\hline \rm{stack} & $N$ & $(n^1,l^1)\times (n^2,l^2)\times
(n^3,l^3)$ & $n_{\Ysymm}$ & $n_{\Yasymm}$ & $b$ & $b'$ & 1 \\
\hline
    $a$&10+2& $(0,-1)\times (1,1)\times (4,1)$ & -3 & 3  & -10 & -18 & 1\\
    $b$&   2& $(-1,1)\times (-2,8)\times (2,1)$ & 38 & 90 &-&- & -8\\
\hline
    1&   4& $(1,0)\times (2,0)\times (1,0)$ & \multicolumn{5}{c|}{$x_A=4x_B, \; 9x_B+4x_C=x_D/2$}\\
\hline
\end{tabular}

\renewcommand{\arraystretch}{1.25}

\begin{tabular}{|c||c|c||c|c|c|c|c|}
\hline
    \rm{model} I.8 & \multicolumn{7}{c|}{$U(5)\times U(1)\times U(1)\times USp(4)$} \\
\hline\hline \rm{stack} & $N$ & $(n^1,l^1)\times (n^2,l^2)\times
(n^3,l^3)$ & $n_{\Ysymm}$ & $n_{\Yasymm}$ & $b$ & $b'$ & 1 \\
\hline
    $a$&10+2& $(0,-1)\times (1,4)\times (1,1)$ & 3 & -3 & 8 & 0 & 4\\
    $b$&   2& $(-1,1)\times (-1,4)\times (4,2)$ & 38 & 90 &- &- & -8\\
\hline
    1&   4& $(1,0)\times (1,0)\times (2,0)$ & \multicolumn{5}{c|}{$x_B=4x_A, \; 9x_B+4x_C=x_D/2$}\\
\hline
\end{tabular}

\begin{tabular}{|c||c|c||c|c|c|c|c|c|}
\hline
    \rm{model} I.9 & \multicolumn{8}{c|}{$U(5)\times U(1)\times U(1)\times USp(2)\times USp(8)$} \\
\hline\hline \rm{stack} & $N$ & $(n^1,l^1)\times (n^2,l^2)\times
(n^3,l^3)$ & $n_{\Ysymm}$ & $n_{\Yasymm}$ & $b$  & $b'$ & 1 & 4 \\
\hline
    $a$&10+2& $(0,-1)\times (1,1)\times (4,1)$ & -3 & 3  & -5 & -21 & 1 & 0 \\
    $b$&   2& $(-1,1)\times (-2,8)\times (3,1)$ & 60 & 132 &- &- & -8& 2\\
\hline
    1&  2& $(1,0)\times (2,0)\times (1,0)$ & \multicolumn{6}{c|}{$x_A=4x_B, \; 19x_B+12x_C=x_D$}\\
    4&  8& $(0,1)\times (0,-2)\times (1,0)$ & \multicolumn{6}{c|}{}\\
\hline
\end{tabular}

\begin{tabular}{|c||c|c||c|c|c|c|c|}
\hline
    \rm{model} I.10 & \multicolumn{7}{c|}{$U(5)\times U(1)\times U(1)\times USp(16)$}\\
\hline\hline \rm{stack} & $N$ & $(n^1,l^1)\times (n^2,l^2)\times
(n^3,l^3)$ & $n_{\Ysymm}$ & $n_{\Yasymm}$ & $b$  & $b'$ & 1 \\
\hline
    $a$&10+2& $(0,-1)\times (1,1)\times (4,1)$ & -3 & 3  & 0 & -24 & 0\\
    $b$&   2& $(-1,1)\times (-2,8)\times (4,1)$ & 82 & 174 &- &- & 2\\
\hline
    4&   16& $(0,1)\times (0,-2)\times (1,0)$ & \multicolumn{5}{c|}{$x_A=4x_B, \; 5x_B+4x_C=x_D/4$}\\
\hline
\end{tabular}

\begin{tabular}{|c||c|c||c|c|c|c|c|c|}
\hline
    \rm{model} II.1.1 & \multicolumn{8}{c|}{$U(5)\times U(1)\times U(1)$} \\
    \hline\hline \rm{stack} & $N$ & $(n^1,l^1)\times (n^2,l^2)\times
(n^3,l^3)$ & $n_{\Ysymm}$ & $n_{\Yasymm}$  & $b$ & $c$ & $b'$ & $c'$ \\
\hline
    $a$&  10& $(0,-1)\times (1,4)\times (1,1)$ & 3 & -3 & 0 & 24  & 0 & 0\\
    $b$&   2& $(-1,3)\times (-1,4)\times (1,1)$ & 6 & 42&- & 0  & - & -96 \\
    $c$&   2& $(-1,0)\times (-1,4)\times (7,1)$ & -27  & 27  &- &- & -& -\\
\hline
     & & \multicolumn{7}{c|}{$x_B=4x_A=4x_C$}\\
\hline
\end{tabular}

\begin{tabular}{|c||c|c||c|c|c|c|c|c|c|}
\hline
    \rm{model} II.1.2 & \multicolumn{9}{c|}{$U(5)\times U(1)\times U(1)\times USp(2)$} \\
    \hline\hline \rm{stack} & $N$ & $(n^1,l^1)\times (n^2,l^2)\times
(n^3,l^3)$ & $n_{\Ysymm}$ & $n_{\Yasymm}$& $b$ & $c$ & $b'$ & $c'$ & 1 \\
\hline
    $a$&  10& $(0,-1)\times (1,4)\times (1,1)$ & 3 & -3  & 0 & 16  & 0 & 0& 4\\
    $b$&   2& $(-1,3)\times (-1,4)\times (1,1)$ & 6 & 42 &- & 0  & - & -72 & -12\\
    $c$&   2& $(-1,0)\times (-1,4)\times (5,1)$ & -19  & 19 &- &- & -& -&0 \\
\hline
    1&2 & $(1,0)\times (1,0)\times (2,0)$ & \multicolumn{7}{c|}{$3x_B=12x_A=8x_C$}\\
\hline
\end{tabular}

\begin{tabular}{|c||c|c||c|c|c|c|c|c|c|}
\hline
    \rm{model} II.1.3 & \multicolumn{9}{c|}{$U(5)\times U(1)\times U(1)\times USp(4)$} \\
    \hline\hline \rm{stack} & $N$ & $(n^1,l^1)\times (n^2,l^2)\times
(n^3,l^3)$ & $n_{\Ysymm}$ & $n_{\Yasymm}$& $b$ & $c$ & $b'$ & $c'$ & 1 \\
\hline
    $a$&  10& $(0,-1)\times (1,4)\times (1,1)$ & 3 & -3  & 0 & 8  & 0 & 0& 4\\
    $b$&   2& $(-1,3)\times (-1,4)\times (1,1)$ & 6 & 42 &- & 0  & - & -48 & -12\\
    $c$&   2& $(-1,0)\times (-1,4)\times (3,1)$ & -11  & 11 &- &- & -& -&0 \\
\hline
    1&4 & $(1,0)\times (1,0)\times (2,0)$ & \multicolumn{7}{c|}{$3x_B=12x_A=4x_C$}\\
\hline
\end{tabular}

\begin{tabular}{|c||c|c||c|c|c|c|c|c|c|}
\hline
    \rm{model} II.1.4 & \multicolumn{9}{c|}{$U(5)\times U(1)\times U(1)\times USp(2)$} \\
    \hline\hline \rm{stack} & $N$ & $(n^1,l^1)\times (n^2,l^2)\times
(n^3,l^3)$ & $n_{\Ysymm}$ & $n_{\Yasymm}$& $b$ & $c$ & $b'$ & $c'$ & 2\\
\hline
    $a$&  10& $(0,-1)\times (1,4)\times (1,1)$ & 3 & -3  & 0 & 18 & 0 & 8 &-1\\
    $b$&   2& $(-1,3)\times (-1,4)\times (1,1)$ & 6 & 42 &- & -18  & - & -72 & -3\\
    $c$&   2& $(-1,0)\times (-1,2)\times (7,1)$ & -13  & 13 &- &- & -& -&0 \\
\hline
    2&2 & $(1,0)\times (0,1)\times (0,-2)$ &\multicolumn{7}{c|}{$7x_B=28x_A=20x_C$}\\
\hline
\end{tabular}

\scriptsize
\renewcommand{\arraystretch}{1.25}
\begin{tabular}{|c||c|c||c|c|c|c|c|c|c|c|}
\hline
    \rm{model} II.1.5 & \multicolumn{10}{c|}{$U(5)\times U(1)\times U(1)\times USp(2)\times
    USp(2)$} \\
    \hline\hline \rm{stack} & $N$ & $(n^1,l^1)\times (n^2,l^2)\times
(n^3,l^3)$ & $n_{\Ysymm}$ & $n_{\Yasymm}$& $b$ & $c$ & $b'$ & $c'$ & 1 & 2\\
\hline
    $a$&  10& $(0,-1)\times (1,4)\times (1,1)$ & 3 & -3  & 0 & 12  & 0 & 6 & 4&-1\\
    $b$&   2& $(-1,3)\times (-1,4)\times (1,1)$ & 6 & 42 &- & -12  & - & -54 & -12&-3\\
    $c$&   2& $(-1,0)\times (-1,2)\times (5,1)$ & -9  & 9 &- &- & -& -&0 &0\\
\hline
    1&2 & $(1,0)\times (1,0)\times (2,0)$ & \multicolumn{8}{c|}{$x_B=4x_A=x_C$}\\
    2&2 & $(1,0)\times (0,1)\times (0,-2)$ &\multicolumn{8}{c|}{}\\
\hline
\end{tabular}

\begin{tabular}{|c||c|c||c|c|c|c|c|c|c|c|}
\hline
    \rm{model} II.1.6 & \multicolumn{10}{c|}{$U(5)\times U(1)\times U(1)\times USp(4)\times
    USp(2)$} \\
    \hline\hline \rm{stack} & $N$ & $(n^1,l^1)\times (n^2,l^2)\times
(n^3,l^3)$ & $n_{\Ysymm}$ & $n_{\Yasymm}$& $b$ & $c$ & $b'$ & $c'$ & 1 & 2\\
\hline
    $a$&  10& $(0,-1)\times (1,4)\times (1,1)$ & 3 & -3  & 0 & 6  & 0 & 4 & 4&-1\\
    $b$&   2& $(-1,3)\times (-1,4)\times (1,1)$ & 6 & 42 &- & -6  & - & 12 & -36 &-3\\
    $c$&   2& $(-1,0)\times (-1,2)\times (3,1)$ & -5  & 5 &- &- & -& -&0 &0\\
\hline
    1&4 & $(1,0)\times (1,0)\times (2,0)$ & \multicolumn{8}{c|}{$3x_B=12x_A=x_C$}\\
    2&2 & $(1,0)\times (0,1)\times (0,-2)$ &\multicolumn{8}{c|}{}\\
\hline
\end{tabular}

\begin{tabular}{|c||c|c||c|c|c|c|c|c|c|c|}
\hline
    \rm{model} II.2.1 & \multicolumn{9}{c|}{$U(5)\times U(1)\times U(1)\times USp(6)$} \\
    \hline\hline \rm{stack} & $N$ & $(n^1,l^1)\times (n^2,l^2)\times
(n^3,l^3)$ & $n_{\Ysymm}$ & $n_{\Yasymm}$  & $b$ & $c$ & $b'$ & $c'$  & 4\\
\hline
    $a$&  10& $(0,-1)\times (1,4)\times (1,1)$ & 3 & -3  & 0& 18 & -2 & 8 &0\\
    $b$&   2& $(-1,3)\times (-1,6)\times (1,1)$ & 10 & 62&- &-36  & - &-96 &1\\
    $c$&   2& $(-1,0)\times (-1,2)\times (7,1)$ & -13 & 13  &- &- & -& -&1\\
\hline
    4&6 & $(0,1)\times (0,-1)\times (2,0)$ &\multicolumn{7}{c|}{$15x_B=60x_A=16x_C$}\\
\hline
\end{tabular}

\begin{tabular}{|c||c|c||c|c|c|c|c|c|c|c|}
\hline
    \rm{model} II.2.2 & \multicolumn{10}{c|}{$U(5)\times U(1)\times U(1)\times USp(2)\times
    USp(6)$} \\
    \hline\hline \rm{stack} & $N$ & $(n^1,l^1)\times (n^2,l^2)\times
(n^3,l^3)$ & $n_{\Ysymm}$ & $n_{\Yasymm}$  & $b$ & $c$ & $b'$ & $c'$ & 1 & 4\\
\hline
    $a$&  10& $(0,-1)\times (1,4)\times (1,1)$ & 3 & -3  & 0& 12 & -2 & 6 & 4&0\\
    $b$&   2& $(-1,3)\times (-1,6)\times (1,1)$ & 10 & 62&- &-24 & - &-72 & -18&1\\
    $c$&   2& $(-1,0)\times (-1,2)\times (5,1)$ & -9 & 9  &- &- & -& -&0 &-1\\
\hline
    1&2 & $(1,0)\times (1,0)\times (2,0)$ & \multicolumn{8}{c|}{$15x_B=60x_A=8x_C$}\\
    4&6 & $(0,1)\times (0,-1)\times (2,0)$ &\multicolumn{8}{c|}{}\\
\hline
\end{tabular}

\begin{tabular}{|c||c|c||c|c|c|c|c|c|c|}
\hline
    \rm{model} II.3.1 & \multicolumn{9}{c|}{$U(5)\times U(1)\times U(1)\times USp(6)$} \\
    \hline\hline \rm{stack} & $N$ & $(n^1,l^1)\times (n^2,l^2)\times
(n^3,l^3)$ & $n_{\Ysymm}$ & $n_{\Yasymm}$ & $b$ & $c$ & $b'$ & $c'$ & 4\\
\hline
    $a$&  10& $(0,-1)\times (1,1)\times (4,1)$ & -3 & 3  & -4& 6 & -6 & -20 & 0\\
    $b$&   2& $(-1,3)\times (-1,3)\times (2,1)$ & 16 & 56&- &12 & - &-96  &1\\
    $c$&   2& $(-1,0)\times (-1,5)\times (6,1)$ & -29 & 29 &- &- & -& -&1\\
\hline
    4&6 & $(0,1)\times (0,-2)\times (1,0)$ &\multicolumn{7}{c|}{$x_A=4x_B=16x_C$}\\
\hline
\end{tabular}

\begin{tabular}{|c||c|c||c|c|c|c|c|c|c|c|}
\hline
    \rm{model} II.3.2 & \multicolumn{10}{c|}{$U(5)\times U(1)\times U(1)\times
    USp(2)\times USp(6)$} \\
    \hline\hline \rm{stack} & $N$ & $(n^1,l^1)\times (n^2,l^2)\times
(n^3,l^3)$ & $n_{\Ysymm}$ & $n_{\Yasymm}$ & $b$ & $c$ & $b'$ & $c'$ & 2 & 4\\
\hline
    $a$&  10& $(0,-1)\times (1,1)\times (4,1)$ & -3 & 3  & -4& 4 & -6 & -10& -4 & 0\\
    $b$&   2& $(-1,3)\times (-1,3)\times (2,1)$ & 16 & 56&- &0 & - &-72 & -6 &1\\
    $c$&   2& $(-1,0)\times (-1,3)\times (6,1)$ & -17 & 17 &- &- & -& -&0&1\\
\hline
    2&2 & $(1,0)\times (0,2)\times (0,-1)$ &\multicolumn{8}{c|}{$x_A=4x_B=8x_C$}\\
    4&6 & $(0,1)\times (0,-2)\times (1,0)$ &\multicolumn{8}{c|}{}\\
\hline
\end{tabular}

\begin{tabular}{|c||c|c||c|c|c|c|c|c|c|c|c|}
\hline
    \rm{model} II.3.3 & \multicolumn{10}{c|}{$U(5)\times U(1)\times U(1)\times USp(2)\times USp(6)$} \\
    \hline\hline \rm{stack} & $N$ & $(n^1,l^1)\times (n^2,l^2)\times
(n^3,l^3)$ & $n_{\Ysymm}$ & $n_{\Yasymm}$ & $b$ & $c$ & $b'$ & $c'$ & 1 & 4\\
\hline
    $a$&  10& $(0,-1)\times (1,1)\times (4,1)$ & -3 & 3  & -4& 0 & -6 & -16& 1& 0\\
    $b$&   2& $(-1,3)\times (-1,3)\times (2,1)$ & 16 & 56&- &6 & - &-72 & -9& 1\\
    $c$&   2& $(-1,0)\times (-1,5)\times (4,1)$ & -19 & 19 &- &- & -& -&0 &1\\
\hline
    1&2 & $(1,0)\times (2,0)\times (1,0)$ & \multicolumn{8}{c|}{$3x_A=12x_B=28x_C$}\\
    4&6 & $(0,1)\times (0,-2)\times (1,0)$ &\multicolumn{8}{c|}{}\\
\hline
\end{tabular}

\scriptsize
\renewcommand{\arraystretch}{1.25}
\begin{tabular}{|c||c|c||c|c|c|c|c|c|c|c|c|}
\hline
    \rm{model} II.3.4 & \multicolumn{11}{c|}{$U(5)\times U(1)\times U(1)\times USp(2)\times
    USp(2)\times USp(6)$} \\
    \hline\hline \rm{stack} & $N$ & $(n^1,l^1)\times (n^2,l^2)\times
(n^3,l^3)$ & $n_{\Ysymm}$ & $n_{\Yasymm}$ & $b$ & $c$ & $b'$ & $c'$ & 1 & 2 & 4\\
\hline
    $a$&  10& $(0,-1)\times (1,1)\times (4,1)$ & -3 & 3  & -4& 0 & -6 & -8& 1& -4 & 0\\
    $b$&   2& $(-1,3)\times (-1,3)\times (2,1)$ & 16 & 56&- &0 & - &-54 & -9& -6 &1\\
    $c$&   2& $(-1,0)\times (-1,3)\times (4,1)$ & -11 & 11 &- &- & -& -&0 &0&-1\\
\hline
    1&2 & $(1,0)\times (2,0)\times (1,0)$ & \multicolumn{9}{c|}{$x_A=4x_B=4x_C$}\\
    2&2 & $(1,0)\times (0,2)\times (0,-1)$ &\multicolumn{9}{c|}{}\\
    4&6 & $(0,1)\times (0,-2)\times (1,0)$ &\multicolumn{9}{c|}{}\\
\hline
\end{tabular}

\begin{tabular}{|c||c|c||c|c|c|c|c|c|c|c|c|}
\hline
    \rm{model} II.3.5 & \multicolumn{10}{c|}{$U(5)\times U(1)\times U(1)\times USp(4)\times USp(6)$} \\
    \hline\hline \rm{stack} & $N$ & $(n^1,l^1)\times (n^2,l^2)\times
(n^3,l^3)$ & $n_{\Ysymm}$ & $n_{\Yasymm}$ & $b$ & $c$ & $b'$ & $c'$ & 1 & 4\\
\hline
    $a$&  10& $(0,-1)\times (1,1)\times (4,1)$ & -3 & 3  & -4& -6 & -6 & -12 & 1 & 0\\
    $b$&   2& $(-1,3)\times (-1,3)\times (2,1)$ & 16 & 56&- &0 & - &-48 & -9 &1\\
    $c$&   2& $(-1,0)\times (-1,5)\times (2,1)$ & -9 & 9 &- &- & -& -&0 &1\\
\hline
    1&4 & $(1,0)\times (2,0)\times (1,0)$ & \multicolumn{8}{c|}{$3x_A=12x_B=8x_C$}\\
    4&6 & $(0,1)\times (0,-2)\times (1,0)$ &\multicolumn{8}{c|}{}\\
\hline
\end{tabular}

\begin{tabular}{|c||c|c||c|c|c|c|c|c|c|}
\hline
    \rm{model} II.4.1 & \multicolumn{9}{c|}{$U(5)\times U(1)\times U(1)\times USp(18)$} \\
    \hline\hline \rm{stack} & $N$ & $(n^1,l^1)\times (n^2,l^2)\times
(n^3,l^3)$ & $n_{\Ysymm}$ & $n_{\Yasymm}$  & $b$ & $c$  & $b'$ & $c'$  & 4\\
\hline
    $a$&  10& $(0,-1)\times (1,1)\times (4,1)$ & -3 & 3  & -6& 4 & -12 & -10&0\\
    $b$&   2& $(-1,3)\times (-1,5)\times (2,1)$ & 30 & 90 &- &-12 & - &-96 &1\\
    $c$&   2& $(-1,0)\times (-1,3)\times (6,1)$ & -17 & 17  &- &- & -& -&1\\
\hline
    4&18 & $(0,1)\times (0,-2)\times (1,0)$ &\multicolumn{7}{c|}{$33x_A=132x_B=16x_C$}\\
\hline
\end{tabular}

\begin{tabular}{|c||c|c||c|c|c|c|c|c|c|c|}
\hline
    \rm{model} II.4.2 & \multicolumn{10}{c|}{$U(5)\times U(1)\times U(1)\times USp(2)\times USp(18)$} \\
    \hline\hline \rm{stack} & $N$ & $(n^1,l^1)\times (n^2,l^2)\times
(n^3,l^3)$ & $n_{\Ysymm}$ & $n_{\Yasymm}$  & $b$ & $c$  & $b'$ & $c'$ & 1 & 4\\
\hline
    $a$&  10& $(0,-1)\times (1,1)\times (4,1)$ & -3 & 3  & -6& 0 & -12 & -8& 1&0\\
    $b$&   2& $(-1,3)\times (-1,5)\times (2,1)$ & 30 & 90 &- &-6& - &-72 & -15&1\\
    $c$&   2& $(-1,0)\times (-1,3)\times (4,1)$ & -11 & 11  &- &- & -& -&0 &1\\
\hline
    1&2 & $(1,0)\times (2,0)\times (1,0)$ & \multicolumn{8}{c|}{$33x_A=132x_B=4x_C$}\\
    4&18 & $(0,1)\times (0,-2)\times (1,0)$ &\multicolumn{8}{c|}{}\\
\hline
\end{tabular}

\begin{tabular}{|c||c|c||c|c|c|c|c|c|}
\hline
    \rm{model} II.5 & \multicolumn{8}{c|}{$U(5)\times U(1)\times U(1)$} \\
    \hline\hline \rm{stack} & $N$ & $(n^1,l^1)\times (n^2,l^2)\times
(n^3,l^3)$ & $n_{\Ysymm}$ & $n_{\Yasymm}$  & $b$ & $c$ & $b'$ & $c'$ \\
\hline
    $a$&  10& $(0,-1)\times (1,7)\times (1,1)$ & 3 & -3  & 5 & 12 & 4 & 3\\
    $b$&   2& $(-1,3)\times (-1,3)\times (3,1)$ & 14 & 40 &- &3& - &-48 \\
    $c$&   2& $(-1,0)\times (-1,5)\times (5,1)$ & -12 & 12  &- &- & -& -\\
\hline
     & \multicolumn{8}{c|}{$9x_B=63x_A=14x_C$}\\
\hline
\end{tabular}

\begin{tabular}{|c||c|c||c|c|c|c|c|c|c|}
\hline
    \rm{model} II.6 & \multicolumn{9}{c|}{$U(5)\times U(1)\times U(1)\times USp(24)$} \\
    \hline\hline \rm{stack} & $N$ & $(n^1,l^1)\times (n^2,l^2)\times
(n^3,l^3)$ & $n_{\Ysymm}$ & $n_{\Yasymm}$  & $b$ & $c$  & $b'$ & $c'$ &  4\\
\hline
    $a$&  10& $(0,-1)\times (1,4)\times (1,1)$ & 3 & -3  & 8 & 16 & 0 & 0& 0\\
    $b$&   2& $(-1,3)\times (-1,4)\times (3,1)$ & 40 & 104&- &0 & - &-96 & 1\\
    $c$&   2& $(-1,0)\times (-1,4)\times (5,1)$ & -19 & 19  &- &- & -& -&1\\
\hline
    4& 24 & $(0,1)\times (0,-1)\times (2,0)$ & \multicolumn{7}{c|}{$3x_B=12x_A=2x_C$}\\
\hline
\end{tabular}

\begin{tabular}{|c||c|c||c|c|c|c|c|c|c|}
\hline
    \rm{model} II.7.1 & \multicolumn{9}{c|}{$U(5)\times U(1)\times U(1)\times USp(4)$} \\
    \hline\hline \rm{stack} & $N$ & $(n^1,l^1)\times (n^2,l^2)\times
(n^3,l^3)$ & $n_{\Ysymm}$ & $n_{\Yasymm}$  & $b$ & $c$  & $b'$ & $c'$  & 4\\
\hline
    $a$&  10& $(0,-1)\times (1,1)\times (4,1)$ & -3 & 3  & -6& 4 & -12 & -10&0\\
    $b$&   2& $(-1,2)\times (-1,5)\times (2,1)$ & 17 & 63 &- &-8 & - &-64 &1\\
    $c$&   2& $(-1,0)\times (-1,3)\times (6,1)$ & -17 & 17  &- &- & -& -&1\\
\hline
    4&4 & $(0,1)\times (0,-2)\times (1,0)$ &\multicolumn{7}{c|}{$11x_A=44x_B=8x_C$}\\
\hline
\end{tabular}

\scriptsize
\renewcommand{\arraystretch}{1.25}
\begin{tabular}{|c||c|c||c|c|c|c|c|c|c|c|}
\hline
    \rm{model} II.7.2 & \multicolumn{10}{c|}{$U(5)\times U(1)\times U(1)\times USp(2)\times USp(4)$} \\
    \hline\hline \rm{stack} & $N$ & $(n^1,l^1)\times (n^2,l^2)\times
(n^3,l^3)$ & $n_{\Ysymm}$ & $n_{\Yasymm}$  & $b$ & $c$  & $b'$ & $c'$ & 1 & 4\\
\hline
    $a$&  10& $(0,-1)\times (1,1)\times (4,1)$ & -3 & 3  & -6& 0 & -12 & -8& 1&0\\
    $b$&   2& $(-1,2)\times (-1,5)\times (2,1)$ & 17 & 63 &- &-4& - &-48 & -10&1\\
    $c$&   2& $(-1,0)\times (-1,3)\times (4,1)$ & -11 & 11  &- &- & -& -&0 &1\\
\hline
    1&2 & $(1,0)\times (2,0)\times (1,0)$ & \multicolumn{8}{c|}{$11x_A=44x_B=2x_C$}\\
    4&4 & $(0,1)\times (0,-2)\times (1,0)$ &\multicolumn{8}{c|}{}\\
\hline
\end{tabular}

\begin{tabular}{|c||c|c||c|c|c|c|c|c|c|c|}
\hline
    \rm{model} III.1.1 & \multicolumn{10}{c|}{$U(5)\times U(1)\times U(4)\times USp(2)\times USp(2)$} \\
    \hline\hline \rm{stack} & $N$ & $(n^1,l^1)\times (n^2,l^2)\times
(n^3,l^3)$ & $n_{\Ysymm}$ & $n_{\Yasymm}$  & $b$ & $c$  & $b'$ & $c'$  & 2 & 4\\
\hline
    $a$&  10& $(0,-1)\times (1,1)\times (4,1)$ & -3 & 3  & -8& 1 & -12 & 0& -4 & 0\\
    $b$&   2& $(-2,3)\times (-1,3)\times (2,1)$ & 47 & 97 &- &-5 & - &-2 & -6 &2\\
    $c$&8& $(1,1)\times (-1,1)\times (-1,0)$ & 0 & 0 &- &- & -& - &1&0\\
\hline
    2& 2& $(1,0)\times (0,2)\times (0,-1)$ &\multicolumn{8}{c|}{$x_A=4x_B=4x_C$}\\
    4& 2& $(0,1)\times (0,-2)\times (1,0)$ &\multicolumn{8}{c|}{}\\
\hline
\end{tabular}

\begin{tabular}{|c||c|c||c|c|c|c|c|c|c|}
\hline
    \rm{model} III.1.2 & \multicolumn{9}{c|}{$U(5)\times U(1)\times U(2)\times USp(2)$} \\
    \hline\hline \rm{stack} & $N$ & $(n^1,l^1)\times (n^2,l^2)\times
(n^3,l^3)$ & $n_{\Ysymm}$ & $n_{\Yasymm}$  & $b$ & $c$  & $b'$ & $c'$ & 2 \\
\hline
    $a$&  10& $(0,-1)\times (1,1)\times (4,1)$ & -3 & 3  & -8& 4 & -12 & -2& -4 \\
    $b$&   2& $(-2,3)\times (-1,3)\times (2,1)$ & 47 & 97 &- &0 & - &-12 & -6 \\
    $c$&4& $(2,1)\times (-1,3)\times (-1,0)$ & -5 & 5 &- &- & -& -&1 \\
\hline
    2& 2& $(1,0)\times (0,2)\times (0,-1)$ &\multicolumn{7}{c|}{$x_A=4x_B=24x_C$}\\
\hline
\end{tabular}

\begin{tabular}{|c||c|c||c|c|c|c|c|c|c|}
\hline
    \rm{model} III.1.3 & \multicolumn{9}{c|}{$U(5)\times U(1)\times U(2)\times USp(2)$} \\
    \hline\hline \rm{stack} & $N$ & $(n^1,l^1)\times (n^2,l^2)\times
(n^3,l^3)$ & $n_{\Ysymm}$ & $n_{\Yasymm}$  & $b$ & $c$  & $b'$ & $c'$ & 2 \\
\hline
    $a$&  10& $(0,-1)\times (1,1)\times (4,1)$ & -3 & 3  & -8& 2 & -12 & 0& -4 \\
    $b$&   2& $(-2,3)\times (-1,3)\times (2,1)$ & 47 & 97 &- &-12 & - &0 & -6 \\
    $c$&4& $(2,3)\times (-1,1)\times (-1,0)$ & -1 & 1 &- &- & -& -&3 \\
\hline
    2& 2& $(1,0)\times (0,2)\times (0,-1)$ &\multicolumn{7}{c|}{$3x_A=12x_B=8x_C$}\\
\hline
\end{tabular}

\begin{tabular}{|c||c|c||c|c|c|c|c|c|c|c|}
\hline
    \rm{model} III.1.4 & \multicolumn{10}{c|}{$U(5)\times U(1)\times U(2)\times USp(2)\times USp(4)$} \\
    \hline\hline \rm{stack} & $N$ & $(n^1,l^1)\times (n^2,l^2)\times
(n^3,l^3)$ & $n_{\Ysymm}$ & $n_{\Yasymm}$  & $b$ & $c$  & $b'$ & $c'$  & 2 & 4\\
\hline
    $a$&  10& $(0,-1)\times (1,1)\times (4,1)$ & -3 & 3  & -8& 2 & -12 & 0& -4 & 0\\
    $b$&   2& $(-2,3)\times (-1,3)\times (2,1)$ & 47 & 97 &- &-8 & - &-8 & -6 &2\\
    $c$&  4 & $(2,1)\times (-1,1)\times (-1,0)$ & -1 & 1 &- &- & -& -&1&0\\
\hline
    2& 2& $(1,0)\times (0,2)\times (0,-1)$ &\multicolumn{8}{c|}{$x_A=4x_B=8x_C$}\\
    4& 4& $(0,1)\times (0,-2)\times (1,0)$ &\multicolumn{8}{c|}{}\\
\hline
\end{tabular}

\begin{tabular}{|c||c|c||c|c|c|c|c|c|c|c|}
\hline
    \rm{model} III.1.5 & \multicolumn{10}{c|}{$U(5)\times U(1)\times U(2)\times USp(2)\times
    USp(2)$} \\
    \hline\hline \rm{stack} & $N$ & $(n^1,l^1)\times (n^2,l^2)\times
(n^3,l^3)$ & $n_{\Ysymm}$ & $n_{\Yasymm}$  & $b$ & $c$  & $b'$ & $c'$ & 1 & 2 \\
\hline
    $a$&  10& $(0,-1)\times (1,1)\times (4,1)$ & -3 & 3  & -8& 2 & -12 & -1& 1& -4 \\
    $b$&   2& $(-2,3)\times (-1,3)\times (2,1)$ & 47 & 97 &- &0 & - &-3 & -9& -6 \\
    $c$&4& $(1,1)\times (-1,3)\times (-1,0)$ & -2 & 2 &- &- & -& -&0 &1\\
\hline
    1&2 & $(1,0)\times (2,0)\times (1,0)$ & \multicolumn{8}{c|}{$x_A=4x_B=12x_C$}\\
    2& 2& $(1,0)\times (0,2)\times (0,-1)$ &\multicolumn{8}{c|}{}\\
\hline
\end{tabular}

\begin{tabular}{|c||c|c||c|c|c|c|c|c|c|c|}
\hline
    \rm{model} III.1.6 & \multicolumn{10}{c|}{$U(5)\times U(1)\times U(2)\times USp(2)\times
    USp(2)$} \\
    \hline\hline \rm{stack} & $N$ & $(n^1,l^1)\times (n^2,l^2)\times
(n^3,l^3)$ & $n_{\Ysymm}$ & $n_{\Yasymm}$  & $b$ & $c$  & $b'$ & $c'$ & 1 & 2 \\
\hline
    $a$&  10& $(0,-1)\times (1,1)\times (4,1)$ & -3 & 3  & -8& 1 & -12 & 0& 1& -4 \\
    $b$&   2& $(-2,3)\times (-1,3)\times (2,1)$ & 47 & 97 &- &-9 & - &6 & -9& -6 \\
    $c$&   4& $(1,3)\times (-1,1)\times (-1,0)$ & 2 & -2 &- &- & -& -&0 &3\\
\hline
    1&2 & $(1,0)\times (2,0)\times (1,0)$ & \multicolumn{8}{c|}{$3x_A=12x_B=4x_C$}\\
    2& 2& $(1,0)\times (0,2)\times (0,-1)$ &\multicolumn{8}{c|}{}\\
\hline
\end{tabular}

\scriptsize
\renewcommand{\arraystretch}{1.25}
\begin{tabular}{|c||c|c||c|c|c|c|c|c|c|c|c|}
\hline
    \rm{model} III.1.7 & \multicolumn{11}{c|}{$U(5)\times U(1)\times U(2)\times USp(2)\times
    USp(2)\times USp(2)$} \\
    \hline\hline \rm{stack} & $N$ & $(n^1,l^1)\times (n^2,l^2)\times
(n^3,l^3)$ & $n_{\Ysymm}$ & $n_{\Yasymm}$  & $b$ & $c$  & $b'$ & $c'$ & 1 & 2 & 4\\
\hline
    $a$&  10& $(0,-1)\times (1,1)\times (4,1)$ & -3 & 3  & -8& 1 & -12 & 0& 1& -4 & 0\\
    $b$&   2& $(-2,3)\times (-1,3)\times (2,1)$ & 47 & 97 &- &-7 & - &2 & -9& -6 &2\\
    $c$&   4& $(1,2)\times (-1,1)\times (-1,0)$ & -1 & 1 &- &- & -& -&0 &2&0\\
\hline
    1&2& $(1,0)\times (2,0)\times (1,0)$ & \multicolumn{9}{c|}{$x_A=4x_B=2x_C$}\\
    2& 2& $(1,0)\times (0,2)\times (0,-1)$ &\multicolumn{9}{c|}{}\\
    4&2& $(0,1)\times (0,-2)\times (1,0)$ &\multicolumn{9}{c|}{}\\
\hline
\end{tabular}

\begin{tabular}{|c||c|c||c|c|c|c|c|c|c|c|c|}
\hline
    \rm{model} III.1.8 & \multicolumn{11}{c|}{$U(5)\times U(1)\times U(2)\times USp(2)\times
    USp(2)\times USp(4)$} \\
    \hline\hline \rm{stack} & $N$ & $(n^1,l^1)\times (n^2,l^2)\times
(n^3,l^3)$ & $n_{\Ysymm}$ & $n_{\Yasymm}$  & $b$ & $c$  & $b'$ & $c'$ & 1 & 2 & 4\\
\hline
    $a$&  10& $(0,-1)\times (1,1)\times (4,1)$ & -3 & 3  & -8& 1 & -12 & 0& 1& -4 & 0\\
    $b$&   2& $(-2,3)\times (-1,3)\times (2,1)$ & 47 & 97 &- &-5 & - &-2 & -9& -6 &2\\
    $c$&$4$& $(1,1)\times (-1,1)\times (-1,0)$ & 0 & 0 &- &- & -& -&0 &1&0\\
\hline
    1&2& $(1,0)\times (2,0)\times (1,0)$ & \multicolumn{9}{c|}{$x_A=4x_B=4x_C$}\\
    2& 2& $(1,0)\times (0,2)\times (0,-1)$ &\multicolumn{9}{c|}{}\\
    4&4& $(0,1)\times (0,-2)\times (1,0)$ &\multicolumn{9}{c|}{}\\
\hline
\end{tabular}

\begin{tabular}{|c||c|c||c|c|c|c|c|c|c|}
\hline
    \rm{model} III.1.9 & \multicolumn{9}{c|}{$U(5)\times U(1)\times U(1)\times USp(2)$} \\
    \hline\hline \rm{stack} & $N$ & $(n^1,l^1)\times (n^2,l^2)\times
(n^3,l^3)$ & $n_{\Ysymm}$ & $n_{\Yasymm}$  & $b$ & $c$  & $b'$ & $c'$ & 2 \\
\hline
    $a$&  10& $(0,-1)\times (1,1)\times (4,1)$ & -3 & 3  & -8& 3 & -12 & 1& -4 \\
    $b$&   2& $(-2,3)\times (-1,3)\times (2,1)$ & 47 & 97 &- &-45 & - &21 & -6 \\
    $c$&   2& $(1,3)\times (-4,2)\times (-1,0)$ & 10 & -10 &- &- & -& -&12 \\
\hline
    2& 2& $(1,0)\times (0,2)\times (0,-1)$ &\multicolumn{7}{c|}{$3x_A=12x_B=2x_C$}\\
\hline
\end{tabular}

\begin{tabular}{|c||c|c||c|c|c|c|c|c|c|}
\hline
    \rm{model} III.1.10 & \multicolumn{9}{c|}{$U(5)\times U(1)\times U(1)\times USp(2)$} \\
    \hline\hline \rm{stack} & $N$ & $(n^1,l^1)\times (n^2,l^2)\times
(n^3,l^3)$ & $n_{\Ysymm}$ & $n_{\Yasymm}$  & $b$ & $c$  & $b'$ & $c'$ & 2 \\
\hline
    $a$&  10& $(0,-1)\times (1,1)\times (4,1)$ & -3 & 3  & -8& 5 & -12 & -1& -4 \\
    $b$&   2& $(-2,3)\times (-1,3)\times (2,1)$ & 47 & 97 &- &-15 & - &-9 & -6 \\
    $c$&   2& $(1,1)\times (-4,6)\times (-1,0)$ & -2& 2 &- &- & -& -&4 \\
\hline
    2& 2& $(1,0)\times (0,2)\times (0,-1)$ &\multicolumn{7}{c|}{$x_A=4x_B=6x_C$}\\
\hline
\end{tabular}

\begin{tabular}{|c||c|c||c|c|c|c|c|c|c|c|}
\hline
    \rm{model} III.1.11 & \multicolumn{10}{c|}{$U(5)\times U(1)\times U(1)\times USp(2)\times USp(4)$} \\
    \hline\hline \rm{stack} & $N$ & $(n^1,l^1)\times (n^2,l^2)\times
(n^3,l^3)$ & $n_{\Ysymm}$ & $n_{\Yasymm}$  & $b$ & $c$  & $b'$ & $c'$  & 2 & 4\\
\hline
    $a$&  10& $(0,-1)\times (1,1)\times (4,1)$ & -3 & 3  & -8& 3 & -12 & 1& -4 & 0\\
    $b$&   2& $(-2,3)\times (-1,3)\times (2,1)$ & 47 & 97 &- &-25 & - &-7 & -6 &2\\
    $c$&  2 & $(1,1)\times (-4,2)\times (-1,0)$ & -2 & 2 &- &- & -& -&4&0\\
\hline
    2& 2& $(1,0)\times (0,2)\times (0,-1)$ &\multicolumn{8}{c|}{$x_A=4x_B=2x_C$}\\
    4& 4& $(0,1)\times (0,-2)\times (1,0)$ &\multicolumn{8}{c|}{}\\
\hline
\end{tabular}

\begin{tabular}{|c||c|c||c|c|c|c|c|c|c|c|c|}
\hline
    \rm{model} III.1.12 & \multicolumn{11}{c|}{$U(5)\times U(1)\times U(1)\times USp(2)\times
    USp(2)\times USp(2)$} \\
    \hline\hline \rm{stack} & $N$ & $(n^1,l^1)\times (n^2,l^2)\times
(n^3,l^3)$ & $n_{\Ysymm}$ & $n_{\Yasymm}$  & $b$ & $c$  & $b'$ & $c'$ & 1 & 2 & 4\\
\hline
    $a$&  10& $(0,-1)\times (1,1)\times (4,1)$ & -3 & 3  & -8& 3 & -12 & -1& 1& -4 & 0\\
    $b$&   2& $(-2,3)\times (-1,3)\times (2,1)$ & 47 & 97 &- &-5 & - & -5 & -9& -6 &2\\
    $c$&   2& $(1,1)\times (-2,4)\times (-1,0)$ & -2 & 2 &- &- & -& -&0 &2&0\\
\hline
    1&2& $(1,0)\times (2,0)\times (1,0)$ & \multicolumn{9}{c|}{$x_A=4x_B=8x_C$}\\
    2& 2& $(1,0)\times (0,2)\times (0,-1)$ &\multicolumn{9}{c|}{}\\
    4&2& $(0,1)\times (0,-2)\times (1,0)$ &\multicolumn{9}{c|}{}\\
\hline
\end{tabular}

\begin{tabular}{|c||c|c||c|c|c|c|c|c|c|c|c|c|}
\hline
    \rm{model} III.2.1 & \multicolumn{9}{c|}{$U(5)\times U(1)\times U(1)\times USp(4)$} \\
    \hline\hline \rm{stack} & $N$ & $(n^1,l^1)\times (n^2,l^2)\times
(n^3,l^3)$ & $n_{\Ysymm}$ & $n_{\Yasymm}$  & $b$ & $c$ & $b'$ & $c'$ & 2 \\
\hline
    $a$&  10& $(0,-1)\times (1,4)\times (1,1)$ & 3 & -3 & 12 & 7 & 8 &1 &-1 \\
    $b$&   2& $(-2,3)\times (-1,2)\times (3,1)$ & 47 & 97&- &5 &- &-5 & -9  \\
    $c$&   2& $(1,1)\times (-1,3)\times (-2,0)$ & -4 & 4 &- &- &- &- & 2\\
\hline
    2&4 & $(1,0)\times (0,1)\times (0,-2)$ & \multicolumn{7}{c|}{$3x_B=12x_A=x_C$}\\
\hline
\end{tabular}

\scriptsize
\renewcommand{\arraystretch}{1.25}
\begin{tabular}{|c||c|c||c|c|c|c|c|c|c|c|c|c|}
\hline
    \rm{model} III.2.2 & \multicolumn{9}{c|}{$U(5)\times U(1)\times U(1)\times USp(4)$} \\
    \hline\hline \rm{stack} & $N$ & $(n^1,l^1)\times (n^2,l^2)\times
(n^3,l^3)$ & $n_{\Ysymm}$ & $n_{\Yasymm}$  & $b$ & $c$ & $b'$ & $c'$ & 2 \\
\hline
    $a$&  10& $(0,-1)\times (1,4)\times (1,1)$ & 3 & -3 & 12 & 5 & 8 &3 &-1 \\
    $b$&   2& $(-2,3)\times (-1,2)\times (3,1)$ & 47 & 97&- &-9 &- &9 & -9  \\
    $c$&   2& $(1,3)\times (-1,1)\times (-2,0)$ & 4 & -4 &- &- &- &- & 6\\
\hline
    2&4 & $(1,0)\times (0,1)\times (0,-2)$ & \multicolumn{7}{c|}{$x_B=4x_A=3x_C$}\\
\hline
\end{tabular}

\begin{tabular}{|c||c|c||c|c|c|c|c|c|c|c|c|c|}
\hline
    \rm{model} III.2.3 & \multicolumn{10}{c|}{$U(5)\times U(1)\times U(1)\times USp(4)\times
    USp(2)$} \\
    \hline\hline \rm{stack} & $N$ & $(n^1,l^1)\times (n^2,l^2)\times
(n^3,l^3)$ & $n_{\Ysymm}$ & $n_{\Yasymm}$  & $b$ & $c$ & $b'$ & $c'$ & 2 & 4\\
\hline
    $a$&  10& $(0,-1)\times (1,4)\times (1,1)$ & 3 & -3 &  12 & 6 & 8 &2&-1 &0\\
    $b$&   2& $(-2,3)\times (-1,2)\times (3,1)$ & 47 & 97&- &0 &- &-4 & -9 &2 \\
    $c$&   2& $(1,1)\times (-1,2)\times (-2,0)$ & -2 & 2 &- &- &- &- & 2& 0\\
\hline
    2&4 & $(1,0)\times (0,1)\times (0,-2)$ & \multicolumn{8}{c|}{$2x_B=8x_A=x_C$}\\
    4&2 & $(0,1)\times (0,-1)\times (2,0)$ &\multicolumn{8}{c|}{}\\
\hline
\end{tabular}

\begin{tabular}{|c||c|c||c|c|c|c|c|c|c|c|c|c|}
\hline
    \rm{model} III.2.4 & \multicolumn{10}{c|}{$U(5)\times U(1)\times U(1)\times USp(4)\times
    USp(2)$} \\
    \hline\hline \rm{stack} & $N$ & $(n^1,l^1)\times (n^2,l^2)\times
(n^3,l^3)$ & $n_{\Ysymm}$ & $n_{\Yasymm}$  & $b$ & $c$ & $b'$ & $c'$ & 2 & 4\\
\hline
    $a$&  10& $(0,-1)\times (1,4)\times (1,1)$ & 3 & -3 &  12 & 5 & 8 &3&-1 &0\\
    $b$&   2& $(-2,3)\times (-1,2)\times (3,1)$ & 47 & 97&- &-7 &- &3 & -9 &2 \\
    $c$&   2& $(1,2)\times (-1,1)\times (-2,0)$ & 2 & -2 &- &- &- &- & 4& 0\\
\hline
    2&4 & $(1,0)\times (0,1)\times (0,-2)$ & \multicolumn{8}{c|}{$x_B=4x_A=2x_C$}\\
    4&2 & $(0,1)\times (0,-1)\times (2,0)$ &\multicolumn{8}{c|}{}\\
\hline
\end{tabular}

\begin{tabular}{|c||c|c||c|c|c|c|c|c|c|c|c|c|}
\hline
    \rm{model} III.2.5 & \multicolumn{10}{c|}{$U(5)\times U(1)\times U(1)\times USp(4)\times
    USp(4)$} \\
    \hline\hline \rm{stack} & $N$ & $(n^1,l^1)\times (n^2,l^2)\times
(n^3,l^3)$ & $n_{\Ysymm}$ & $n_{\Yasymm}$  & $b$ & $c$ & $b'$ & $c'$ & 2 & 4\\
\hline
    $a$&  10& $(0,-1)\times (1,4)\times (1,1)$ & 3 & -3 &  12 & 5 & 8 &3&-1 &0\\
    $b$&   2& $(-2,3)\times (-1,2)\times (3,1)$ & 47 & 97&- &-5 &- &-3 & -9 &2 \\
    $c$&   2& $(1,1)\times (-1,1)\times (-2,0)$ & 0 & 0 &- &- &- &- & 2& 0\\
\hline
    2&4 & $(1,0)\times (0,1)\times (0,-2)$ & \multicolumn{8}{c|}{$x_B=4x_A=x_C$}\\
    4&4 & $(0,1)\times (0,-1)\times (2,0)$ &\multicolumn{8}{c|}{}\\
\hline
\end{tabular}

\begin{tabular}{|c||c|c||c|c|c|c|c|c|c|}
\hline
    \rm{model} III.3.1 & \multicolumn{9}{c|}{$U(5)\times U(1)\times U(1)\times USp(2)$} \\
    \hline\hline \rm{stack} & $N$ & $(n^1,l^1)\times (n^2,l^2)\times
(n^3,l^3)$ & $n_{\Ysymm}$ & $n_{\Yasymm}$  & $b$ & $c$ & $b'$ & $c'$ & 2 \\
\hline
    $a$&  10& $(0,-1)\times (1,1)\times (4,1)$ & -3 & 3 & -4 & 6 & -14 & -4 &-4\\
    $b$&   2& $(-2,3)\times (-1,3)\times (3,1)$ & 74 & 142 &- &12 & - &0 & -9\\
    $c$&   2& $(2,3)\times (-1,5)\times (-1,0)$ & -7 & 7 &- &- & -& -&3 \\
\hline
    2&2 & $(1,0)\times (0,2)\times (0,-1)$ & \multicolumn{7}{c|}{$3x_A=12x_B=40x_C$}\\
\hline
\end{tabular}

\begin{tabular}{|c||c|c||c|c|c|c|c|c|c|}
\hline
    \rm{model} III.3.2 & \multicolumn{9}{c|}{$U(5)\times U(1)\times U(1)\times USp(2)$} \\
    \hline\hline \rm{stack} & $N$ & $(n^1,l^1)\times (n^2,l^2)\times
(n^3,l^3)$ & $n_{\Ysymm}$ & $n_{\Yasymm}$  & $b$ & $c$ & $b'$ & $c'$ & 2 \\
\hline
    $a$&  10& $(0,-1)\times (1,1)\times (4,1)$ & -3 & 3 & -4 & 2 & -14 & -4 &-4\\
    $b$&   2& $(-2,3)\times (-1,3)\times (3,1)$ & 74 & 142 &- &48 & - &0 & -9\\
    $c$&   2& $(2,5)\times (-1,3)\times (-1,0)$ & -1 & 1 &- &- & -& -&5 \\
\hline
    2&2 & $(1,0)\times (0,2)\times (0,-1)$ & \multicolumn{7}{c|}{$5x_A=20x_B=24x_C$}\\
\hline
\end{tabular}

\begin{tabular}{|c||c|c||c|c|c|c|c|c|c|}
\hline
    \rm{model} III.3.3 & \multicolumn{9}{c|}{$U(5)\times U(1)\times U(1)\times USp(2)$} \\
    \hline\hline \rm{stack} & $N$ & $(n^1,l^1)\times (n^2,l^2)\times
(n^3,l^3)$ & $n_{\Ysymm}$ & $n_{\Yasymm}$  & $b$ & $c$ & $b'$ & $c'$ & 2 \\
\hline
    $a$&  10& $(0,-1)\times (1,1)\times (4,1)$ & -3 & 3 & -4 & 16 & -14 & -14 &-4\\
    $b$&   2& $(-2,3)\times (-1,3)\times (3,1)$ & 74 & 142 &- &48 & - &-36 & -9\\
    $c$&   2& $(2,1)\times (-1,15)\times (-1,0)$ & -29 & 29 &- &- & -& -&1 \\
\hline
    2&2 & $(1,0)\times (0,2)\times (0,-1)$ & \multicolumn{7}{c|}{$x_A=4x_B=120x_C$}\\
\hline
\end{tabular}

\scriptsize
\renewcommand{\arraystretch}{1.25}
\begin{tabular}{|c||c|c||c|c|c|c|c|c|c|}
\hline
    \rm{model} III.3.4 & \multicolumn{9}{c|}{$U(5)\times U(1)\times U(1)\times USp(2)$} \\
    \hline\hline \rm{stack} & $N$ & $(n^1,l^1)\times (n^2,l^2)\times
(n^3,l^3)$ & $n_{\Ysymm}$ & $n_{\Yasymm}$  & $b$ & $c$ & $b'$ & $c'$ & 2 \\
\hline
    $a$&  10& $(0,-1)\times (1,1)\times (4,1)$ & -3 & 3 & -4 & 2 & -14 & 0 &-4\\
    $b$&   2& $(-2,3)\times (-1,3)\times (3,1)$ & 74 & 142 &- &-36 & - &48 & -9\\
    $c$&   2& $(2,15)\times (-1,1)\times (-1,0)$ & 13 & -13 &- &- & -& -&15 \\
\hline
    2&2 & $(1,0)\times (0,2)\times (0,-1)$ & \multicolumn{7}{c|}{$15x_A=60x_B=4x_C$}\\
\hline
\end{tabular}

\begin{tabular}{|c||c|c||c|c|c|c|c|c|c|c|}
\hline
    \rm{model} III.3.5 & \multicolumn{10}{c|}{$U(5)\times U(1)\times U(1)\times USp(2)\times
    USp(14)$} \\
    \hline\hline \rm{stack} & $N$ & $(n^1,l^1)\times (n^2,l^2)\times
(n^3,l^3)$ & $n_{\Ysymm}$ & $n_{\Yasymm}$  & $b$ & $c$ & $b'$ & $c'$ & 2 & 4\\
\hline
    $a$&  10& $(0,-1)\times (1,1)\times (4,1)$ & 3 & -3 & -4& 2 & -14 & 0& -4&0\\
    $b$&   2& $(-2,3)\times (-1,3)\times (3,1)$ & 74 & 142 &- &-8 & - &-8 & -9&2\\
    $c$&   2& $(2,1)\times (-1,1)\times (-1,0)$ & -1 & 1 &- &- & -& -&1 &0\\
\hline
    2&2 & $(1,0)\times (0,2)\times (0,-1)$ & \multicolumn{8}{c|}{$x_A=4x_B=8x_C$}\\
    4&14 & $(0,1)\times (0,-2)\times (1,0)$ &\multicolumn{8}{c|}{}\\
\hline
\end{tabular}

\begin{tabular}{|c||c|c||c|c|c|c|c|c|c|c|}
\hline
    \rm{model} III.3.6 & \multicolumn{10}{c|}{$U(5)\times U(1)\times U(1)\times USp(2)\times USp(6)$} \\
    \hline\hline \rm{stack} & $N$ & $(n^1,l^1)\times (n^2,l^2)\times
(n^3,l^3)$ & $n_{\Ysymm}$ & $n_{\Yasymm}$  & $b$ & $c$ & $b'$ & $c'$ & 2 & 4\\
\hline
    $a$&  10& $(0,-1)\times (1,1)\times (4,1)$ & 3 & -3 & -4& 4 & -14 & -2& -4&0\\
    $b$&   2& $(-2,3)\times (-1,3)\times (3,1)$ & 74 & 142 &- &0 & - &0 & -9&2\\
    $c$&   2& $(2,1)\times (-1,1)\times (-1,0)$ & -3 & 3 &- &- & -& -&3 &0\\
\hline
    2&2 & $(1,0)\times (0,2)\times (0,-1)$ & \multicolumn{8}{c|}{$x_A=4x_B=8x_C$}\\
    4&6 & $(0,1)\times (0,-2)\times (1,0)$ &\multicolumn{8}{c|}{}\\
\hline
\end{tabular}

\begin{tabular}{|c||c|c||c|c|c|c|c|c|c|c|}
\hline
    \rm{model} III.3.7 & \multicolumn{10}{c|}{$U(5)\times U(1)\times U(1)\times USp(2)\times
    USp(12)$} \\
    \hline\hline \rm{stack} & $N$ & $(n^1,l^1)\times (n^2,l^2)\times
(n^3,l^3)$ & $n_{\Ysymm}$ & $n_{\Yasymm}$  & $b$ & $c$ & $b'$ & $c'$ & 2 & 4\\
\hline
    $a$&  10& $(0,-1)\times (1,1)\times (4,1)$ & 3 & -3 & -4& 4 & -14 & -2& -4&0\\
    $b$&   2& $(-2,3)\times (-1,3)\times (3,1)$ & 74 & 142 &- &0 & - &-12 & -9&2\\
    $c$&   2& $(2,1)\times (-1,3)\times (-1,0)$ & -5 & 5 &- &- & -& -&1 &0\\
\hline
    2&2 & $(1,0)\times (0,2)\times (0,-1)$ & \multicolumn{8}{c|}{$x_A=4x_B=24x_C$}\\
    4&12 & $(0,1)\times (0,-2)\times (1,0)$ &\multicolumn{8}{c|}{}\\
\hline
\end{tabular}

\begin{tabular}{|c||c|c||c|c|c|c|c|c|c|c|}
\hline
    \rm{model} III.3.8 & \multicolumn{10}{c|}{$U(5)\times U(1)\times U(1)\times USp(2)\times
    USp(12)$} \\
    \hline\hline \rm{stack} & $N$ & $(n^1,l^1)\times (n^2,l^2)\times
(n^3,l^3)$ & $n_{\Ysymm}$ & $n_{\Yasymm}$  & $b$ & $c$ & $b'$ & $c'$ & 2 & 4\\
\hline
    $a$&  10& $(0,-1)\times (1,1)\times (4,1)$ & 3 & -3 & -4& 2 & -14 & 0& -4&0\\
    $b$&   2& $(-2,3)\times (-1,3)\times (3,1)$ & 74 & 142 &- &-12 & - &0 & -9&2\\
    $c$&   2& $(2,3)\times (-1,1)\times (-1,0)$ & 1 & -1 &- &- & -& -&3 &0\\
\hline
    2&2 & $(1,0)\times (0,2)\times (0,-1)$ & \multicolumn{8}{c|}{$3x_A=12x_B=8x_C$}\\
    4&12 & $(0,1)\times (0,-2)\times (1,0)$ &\multicolumn{8}{c|}{}\\
\hline
\end{tabular}

\begin{tabular}{|c||c|c||c|c|c|c|c|c|c|c|}
\hline
    \rm{model} III.3.9 & \multicolumn{10}{c|}{$U(5)\times U(1)\times U(1)\times USp(2)\times
    USp(10)$} \\
    \hline\hline \rm{stack} & $N$ & $(n^1,l^1)\times (n^2,l^2)\times
(n^3,l^3)$ & $n_{\Ysymm}$ & $n_{\Yasymm}$  & $b$ & $c$ & $b'$ & $c'$ & 2 & 4\\
\hline
    $a$&  10& $(0,-1)\times (1,1)\times (4,1)$ & 3 & -3 & -4& 6 & -14 & -4& -4&0\\
    $b$&   2& $(-2,3)\times (-1,3)\times (3,1)$ & 74 & 142 &- &8 & - &-16 & -9&2\\
    $c$&   2& $(2,1)\times (-1,5)\times (-1,0)$ & -9 & 9 &- &- & -& -&1 &0\\
\hline
    2&2 & $(1,0)\times (0,2)\times (0,-1)$ & \multicolumn{8}{c|}{$x_A=4x_B=40x_C$}\\
    4&10 & $(0,1)\times (0,-2)\times (1,0)$ &\multicolumn{8}{c|}{}\\
\hline
\end{tabular}

\begin{tabular}{|c||c|c||c|c|c|c|c|c|c|c|}
\hline
    \rm{model} III.3.10 & \multicolumn{10}{c|}{$U(5)\times U(1)\times U(1)\times USp(2)\times
    USp(10)$} \\
    \hline\hline \rm{stack} & $N$ & $(n^1,l^1)\times (n^2,l^2)\times
(n^3,l^3)$ & $n_{\Ysymm}$ & $n_{\Yasymm}$  & $b$ & $c$ & $b'$ & $c'$ & 2 & 4\\
\hline
    $a$&  10& $(0,-1)\times (1,1)\times (4,1)$ & 3 & -3 & -4& 2 & -14 & 0& -4&0\\
    $b$&   2& $(-2,3)\times (-1,3)\times (3,1)$ & 74 & 142 &- &-16 & - &8 & -9&2\\
    $c$&   2& $(2,5)\times (-1,1)\times (-1,0)$ & 3 & -3 &- &- & -& -&5 &0\\
\hline
    2&2 & $(1,0)\times (0,2)\times (0,-1)$ & \multicolumn{8}{c|}{$5x_A=20x_B=40x_C$}\\
    4&10 & $(0,1)\times (0,-2)\times (1,0)$ &\multicolumn{8}{c|}{}\\
\hline
\end{tabular}

\scriptsize
\renewcommand{\arraystretch}{1.25}
\begin{tabular}{|c||c|c||c|c|c|c|c|c|c|c|}
\hline
    \rm{model} III.3.11 & \multicolumn{10}{c|}{$U(5)\times U(1)\times U(1)\times USp(2)\times
    USp(8)$} \\
    \hline\hline \rm{stack} & $N$ & $(n^1,l^1)\times (n^2,l^2)\times
(n^3,l^3)$ & $n_{\Ysymm}$ & $n_{\Yasymm}$  & $b$ & $c$ & $b'$ & $c'$ & 2 & 4\\
\hline
    $a$&  10& $(0,-1)\times (1,1)\times (4,1)$ & 3 & -3 & -4& 8 & -14 & -6& -4&0\\
    $b$&   2& $(-2,3)\times (-1,3)\times (3,1)$ & 74 & 142 &- &16 & - &-20 & -9&2\\
    $c$&   2& $(2,1)\times (-1,7)\times (-1,0)$ & -13 & 13 &- &- & -& -&1 &0\\
\hline
    2&2 & $(1,0)\times (0,2)\times (0,-1)$ & \multicolumn{8}{c|}{$x_A=4x_B=56x_C$}\\
    4&8 & $(0,1)\times (0,-2)\times (1,0)$ &\multicolumn{8}{c|}{}\\
\hline
\end{tabular}

\begin{tabular}{|c||c|c||c|c|c|c|c|c|c|c|}
\hline
    \rm{model} III.3.12 & \multicolumn{10}{c|}{$U(5)\times U(1)\times U(1)\times USp(2)\times
    USp(8)$} \\
    \hline\hline \rm{stack} & $N$ & $(n^1,l^1)\times (n^2,l^2)\times
(n^3,l^3)$ & $n_{\Ysymm}$ & $n_{\Yasymm}$  & $b$ & $c$ & $b'$ & $c'$ & 2 & 4\\
\hline
    $a$&  10& $(0,-1)\times (1,1)\times (4,1)$ & 3 & -3 & -4& 2 & -14 & 0& -4&0\\
    $b$&   2& $(-2,3)\times (-1,3)\times (3,1)$ & 74 & 142 &- &-20 & - &16 & -9&2\\
    $c$&   2& $(2,7)\times (-1,1)\times (-1,0)$ & 5 & -5 &- &- & -& -&7 &0\\
\hline
    2&2 & $(1,0)\times (0,2)\times (0,-1)$ & \multicolumn{8}{c|}{$7x_A=28x_B=8x_C$}\\
    4&8 & $(0,1)\times (0,-2)\times (1,0)$ &\multicolumn{8}{c|}{}\\
\hline
\end{tabular}

\begin{tabular}{|c||c|c||c|c|c|c|c|c|c|c|}
\hline
    \rm{model} III.3.13 & \multicolumn{10}{c|}{$U(5)\times U(1)\times U(1)\times USp(2)\times
    USp(6)$} \\
    \hline\hline \rm{stack} & $N$ & $(n^1,l^1)\times (n^2,l^2)\times
(n^3,l^3)$ & $n_{\Ysymm}$ & $n_{\Yasymm}$  & $b$ & $c$ & $b'$ & $c'$ & 2 & 4\\
\hline
    $a$&  10& $(0,-1)\times (1,1)\times (4,1)$ & 3 & -3 & -4& 10 & -14 & -8& -4&0\\
    $b$&   2& $(-2,3)\times (-1,3)\times (3,1)$ & 74 & 142 &- &24 & - &-24 & -9&2\\
    $c$&   2& $(2,1)\times (-1,9)\times (-1,0)$ & -17 & 17 &- &- & -& -&1 &0\\
\hline
    2&2 & $(1,0)\times (0,2)\times (0,-1)$ & \multicolumn{8}{c|}{$x_A=4x_B=72x_C$}\\
    4&6 & $(0,1)\times (0,-2)\times (1,0)$ &\multicolumn{8}{c|}{}\\
\hline
\end{tabular}

\begin{tabular}{|c||c|c||c|c|c|c|c|c|c|c|}
\hline
    \rm{model} III.3.14 & \multicolumn{10}{c|}{$U(5)\times U(1)\times U(1)\times USp(2)\times
    USp(6)$} \\
    \hline\hline \rm{stack} & $N$ & $(n^1,l^1)\times (n^2,l^2)\times
(n^3,l^3)$ & $n_{\Ysymm}$ & $n_{\Yasymm}$  & $b$ & $c$ & $b'$ & $c'$ & 2 & 4\\
\hline
    $a$&  10& $(0,-1)\times (1,1)\times (4,1)$ & 3 & -3 & -4& 2 & -14 & 0& -4&0\\
    $b$&   2& $(-2,3)\times (-1,3)\times (3,1)$ & 74 & 142 &- &-24 & - &24 & -9&2\\
    $c$&   2& $(2,9)\times (-1,1)\times (-1,0)$ & 7 & -7 &- &- & -& -&9 &0\\
\hline
    2&2 & $(1,0)\times (0,2)\times (0,-1)$ & \multicolumn{8}{c|}{$9x_A=36x_B=8x_C$}\\
    4&6 & $(0,1)\times (0,-2)\times (1,0)$ &\multicolumn{8}{c|}{}\\
\hline
\end{tabular}

\begin{tabular}{|c||c|c||c|c|c|c|c|c|c|c|}
\hline
    \rm{model} III.3.15 & \multicolumn{10}{c|}{$U(5)\times U(1)\times U(1)\times USp(2)\times
    USp(4)$} \\
    \hline\hline \rm{stack} & $N$ & $(n^1,l^1)\times (n^2,l^2)\times
(n^3,l^3)$ & $n_{\Ysymm}$ & $n_{\Yasymm}$  & $b$ & $c$ & $b'$ & $c'$ & 2 & 4\\
\hline
    $a$&  10& $(0,-1)\times (1,1)\times (4,1)$ & 3 & -3 & -4& 12 & -14 & -10& -4&0\\
    $b$&   2& $(-2,3)\times (-1,3)\times (3,1)$ & 74 & 142 &- &32 & - &-28 & -9&2\\
    $c$&   2& $(2,1)\times (-1,11)\times (-1,0)$ & -21 & 21 &- &- & -& -&1 &0\\
\hline
    2&2 & $(1,0)\times (0,2)\times (0,-1)$ & \multicolumn{8}{c|}{$x_A=4x_B=88x_C$}\\
    4&4 & $(0,1)\times (0,-2)\times (1,0)$ &\multicolumn{8}{c|}{}\\
\hline
\end{tabular}

\begin{tabular}{|c||c|c||c|c|c|c|c|c|c|c|}
\hline
    \rm{model} III.3.16 & \multicolumn{10}{c|}{$U(5)\times U(1)\times U(1)\times USp(2)\times
    USp(4)$} \\
    \hline\hline \rm{stack} & $N$ & $(n^1,l^1)\times (n^2,l^2)\times
(n^3,l^3)$ & $n_{\Ysymm}$ & $n_{\Yasymm}$  & $b$ & $c$ & $b'$ & $c'$ & 2 & 4\\
\hline
    $a$&  10& $(0,-1)\times (1,1)\times (4,1)$ & 3 & -3 & -4& 2 & -14 & 0& -4&0\\
    $b$&   2& $(-2,3)\times (-1,3)\times (3,1)$ & 74 & 142 &- &-28 & - &32 & -9&2\\
    $c$&   2& $(2,11)\times (-1,1)\times (-1,0)$ & 9 & -9 &- &- & -& -&11 &0\\
\hline
    2&2 & $(1,0)\times (0,2)\times (0,-1)$ & \multicolumn{8}{c|}{$11x_A=44x_B=8x_C$}\\
    4&4 & $(0,1)\times (0,-2)\times (1,0)$ &\multicolumn{8}{c|}{}\\
\hline
\end{tabular}

\begin{tabular}{|c||c|c||c|c|c|c|c|c|c|c|}
\hline
    \rm{model} III.3.17 & \multicolumn{10}{c|}{$U(5)\times U(1)\times U(1)\times USp(2)\times
    USp(2)$} \\
    \hline\hline \rm{stack} & $N$ & $(n^1,l^1)\times (n^2,l^2)\times
(n^3,l^3)$ & $n_{\Ysymm}$ & $n_{\Yasymm}$  & $b$ & $c$ & $b'$ & $c'$ & 2 & 4\\
\hline
    $a$&  10& $(0,-1)\times (1,1)\times (4,1)$ & 3 & -3 & -4& 14 & -14 & -12 & -4&0\\
    $b$&   2& $(-2,3)\times (-1,3)\times (3,1)$ & 74 & 142 &- &40 & - &-32 & -9&2\\
    $c$&   2& $(2,1)\times (-1,13)\times (-1,0)$ & -25 & 25 &- &- & -& -&1 &0\\
\hline
    2&2 & $(1,0)\times (0,2)\times (0,-1)$ & \multicolumn{8}{c|}{$x_A=4x_B=104x_C$}\\
    4&2 & $(0,1)\times (0,-2)\times (1,0)$ &\multicolumn{8}{c|}{}\\
\hline
\end{tabular}

\scriptsize
\renewcommand{\arraystretch}{1.25}
\begin{tabular}{|c||c|c||c|c|c|c|c|c|c|c|}
\hline
    \rm{model} III.3.18 & \multicolumn{10}{c|}{$U(5)\times U(1)\times U(1)\times USp(2)\times
    USp(2)$} \\
    \hline\hline \rm{stack} & $N$ & $(n^1,l^1)\times (n^2,l^2)\times
(n^3,l^3)$ & $n_{\Ysymm}$ & $n_{\Yasymm}$  & $b$ & $c$ & $b'$ & $c'$ & 2 & 4\\
\hline
    $a$&  10& $(0,-1)\times (1,1)\times (4,1)$ & 3 & -3 & -4& 2 & -14 & 0 & -4&0\\
    $b$&   2& $(-2,3)\times (-1,3)\times (3,1)$ & 74 & 142 &- &-32 & - &40 & -9&2\\
    $c$&   2& $(2,13)\times (-1,1)\times (-1,0)$ & 11 & -11 &- &- & -& -&13 &0\\
\hline
    2&2 & $(1,0)\times (0,2)\times (0,-1)$ & \multicolumn{8}{c|}{$13x_A=52x_B=8x_C$}\\
    4&2 & $(0,1)\times (0,-2)\times (1,0)$ &\multicolumn{8}{c|}{}\\
\hline
\end{tabular}

\begin{tabular}{|c||c|c||c|c|c|c|c|c|c|c|c|}
\hline
    \rm{model} & \multicolumn{9}{c|}{$U(5)\times U(1)\times U(1)\times USp(N^{(4)})$} \\
    III.4.1-56 & \multicolumn{9}{c|}{} \\
    \hline\hline \rm{stack} & $N$ & $(n^1,l^1)\times (n^2,l^2)\times
(n^3,l^3)$ & $n_{\Ysymm}$ & $n_{\Yasymm}$  & $b$ & $c$  & $b'$ & $c'$ & 4\\
\hline
    $a$&  10& $(0,-1)\times (1,4)\times (1,1)$ & 3 & -3 & 16 & $4+l_c^2$& 0 & $4-l_c^2$& 0 \\
    $b$&   2& $(-2,3)\times (-1,4)\times (3,1)$ & 101 & 187&- &$36-2N^{(4)}$ &- &$36-2N^{(4)}$ & 2 \\
     &    &  &  & & &$+3l_c^2-8l_c^1$ &  &$-3l_c^2+8l_c^1$ &  \\
    $c$&   2& $(1,l_c^1)\times (-1,l_c^2)\times (-2,0)$ & $2l_c^1$ & $-2l_c^1$  &- &- &- &- & 0 \\
     &    &  & $-2l_c^2$  &$+2l_c^2$ &  &  &  & &  \\
\hline
    4&$N^{(4)}$ & $(0,1)\times (0,-1)\times (2,0)$ & \multicolumn{7}{c|}{$x_B=4x_A=l_c^2 x_C/l_c^1,\;\; l_c^1l_c^2=24-N^{(4)}$}\\
\hline
\end{tabular}

\begin{tabular}{|c||c|c||c|c|c|c|c|c|c|c|c|}
\hline
    \rm{model} & \multicolumn{9}{c|}{$U(5)\times U(2)\times U(1)\times USp(N^{(4)})$} \\
    III.5.1-13 & \multicolumn{9}{c|}{} \\
    \hline\hline \rm{stack} & $N$ & $(n^1,l^1)\times (n^2,l^2)\times
(n^3,l^3)$ & $n_{\Ysymm}$ & $n_{\Yasymm}$  & $b$ & $c$  & $b'$ & $c'$ & 4\\
\hline
    $a$&  10& $(0,-1)\times (1,4)\times (1,1)$ & 3 & -3 & 8 & $4+l_c^2$& 0 & $4-l_c^2$& 0 \\
    $b$&   4& $(-1,1)\times (-1,4)\times (3,1)$ & 6 & 42&- &$4-N^{(4)}$ &- &$4-N^{(4)}$ & 1 \\
     &    &  &  & & &$+l_c^2-4l_c^1$ &  &$-l_c^2+4l_c^1$ &  \\
    $c$&   2& $(1,l_c^1)\times (-1,l_c^2)\times (-2,0)$ & $2l_c^1$ & $-2l_c^1$  &- &- &- &- & 0 \\
     &    &  & $-2l_c^2$  &$+2l_c^2$ &  &  &  & &  \\
\hline
    4&$N^{(4)}$ & $(0,1)\times (0,-1)\times (2,0)$ & \multicolumn{7}{c|}{$x_B=4x_A=l_c^2 x_C/l_c^1,\;\; l_c^1l_c^2=8-N^{(4)}$}\\
\hline
\end{tabular}

\begin{tabular}{|c||c|c||c|c|c|c|c|c|c|c|}
\hline
    \rm{model} III.6.1 & \multicolumn{9}{c|}{$U(5)\times U(2)\times U(2)\times USp(2)$}\\
    \hline\hline \rm{stack} & $N$ & $(n^1,l^1)\times (n^2,l^2)\times
(n^3,l^3)$ & $n_{\Ysymm}$ & $n_{\Yasymm}$ & $b$ & $c$ & $b'$ & $c'$ & 2 \\
\hline
    $a$&  10& $(0,-1)\times (1,1)\times (4,1)$ & -3 & 3 & -2& 1 & -7 & 0& -4\\
    $b$&   4& $(-1,1)\times (-1,3)\times (3,1)$ & 4 & 32 &- &-2 & - &0 & -3\\
    $c$&   4& $(1,1)\times (-1,1)\times (-1,0)$ & 0 & 0  &- &- & -& -&1 \\
\hline
    2&2 & $(1,0)\times (0,2)\times (0,-1)$ & \multicolumn{7}{c|}{$x_A=4x_B=4x_C$}\\
\hline
\end{tabular}

\begin{tabular}{|c||c|c||c|c|c|c|c|c|c|c|}
\hline
    \rm{model} III.6.2 & \multicolumn{9}{c|}{$U(5)\times U(2)\times U(1)\times USp(2)$}\\
    \hline\hline \rm{stack} & $N$ & $(n^1,l^1)\times (n^2,l^2)\times
(n^3,l^3)$ & $n_{\Ysymm}$ & $n_{\Yasymm}$ & $b$ & $c$ & $b'$ & $c'$ & 2 \\
\hline
    $a$&  10& $(0,-1)\times (1,4)\times (1,1)$ & 3 & -3 & 7& 5 & 2 & 3& -1\\
    $b$&   4& $(-1,1)\times (-1,3)\times (3,1)$ & 4 & 32 &- &-4 & - &0 & -3\\
    $c$&   2& $(1,1)\times (-1,1)\times (-2,0)$ & 0 & 0  &- &- & -& -&2 \\
\hline
    2&2 & $(1,0)\times (0,1)\times (0,-2)$ & \multicolumn{7}{c|}{$x_B=4x_A=x_C$}\\
\hline
\end{tabular}

\begin{tabular}{|c||c|c||c|c|c|c|c|c|c|c|c|c|}
\hline
    \rm{model} & \multicolumn{10}{c|}{$U(5)\times U(1)\times U(1)\times USp(2)\times USp(N^{(4)})$} \\
    III.7.1-5 & \multicolumn{10}{c|}{} \\
    \hline\hline \rm{stack} & $N$ & $(n^1,l^1)\times (n^2,l^2)\times
(n^3,l^3)$ & $n_{\Ysymm}$ & $n_{\Yasymm}$  & $b$ & $c$  & $b'$ & $c'$ & 2 & 4\\
\hline
    $a$&  10& $(0,-1)\times (1,1)\times (4,1)$ & -3 & 3 & -2 & $(1+l_c^2)/2$& -15 & $(1-l_c^2)/2$&-4 & 0 \\
    $b$&   2& $(-1,1)\times (-1,3)\times (7,2)$ & 52 & 116&- &$2-N^{(4)}$ &- &$2-N^{(4)}$ & -7& 2 \\
     &    &  &  & & &$+l_c^2-3l_c^1$ &  &$-l_c^2+3l_c^1$ & & \\
    $c$&   2& $(1,l_c^1)\times (-1,l_c^2)\times (-1,0)$ & $l_c^1$ & $-l_c^1$  &- &- &- &- & $l_c^1$ &0\\
     &    &  & $-l_c^2$  &$+l_c^2$ &  &  &  & & & \\
\hline
    2&2 & $(1,0)\times (0,2)\times (0,-1)$ & \multicolumn{8}{c|}{$x_A=4x_B=4l_c^2 x_C/l_c^1,\;\; l_c^1l_c^2=5-N^{(4)}$}\\
    4&$N^{(4)}$ & $(0,1)\times (0,-2)\times (1,0)$ & \multicolumn{8}{c|}{}\\
\hline
\end{tabular}

\scriptsize
\renewcommand{\arraystretch}{1.25}
\begin{tabular}{|c||c|c||c|c|c|c|c|c|c|c|c|}
\hline
    \rm{model} & \multicolumn{9}{c|}{$U(5)\times U(1)\times U(2)\times USp(N^{(4)})$} \\
    III.8.1-6 & \multicolumn{9}{c|}{} \\
    \hline\hline \rm{stack} & $N$ & $(n^1,l^1)\times (n^2,l^2)\times
(n^3,l^3)$ & $n_{\Ysymm}$ & $n_{\Yasymm}$  & $b$ & $c$  & $b'$ & $c'$  & 4\\
\hline
    $a$&  10& $(0,-1)\times (1,1)\times (4,1)$ & -3 & 3 & -5 & $(1+l_c^2)/2$& -21 & $(1-l_c^2)/2$ & 0 \\
    $b$&   2& $(-1,1)\times (-2,8)\times (3,1)$ & 60 & 132&- &$-N^{(4)}/2$ &- &$-N^{(4)}/2$ & 2 \\
     &    &  &  & & &$+l_c^2-4l_c^1$ &  &$-l_c^2+4l_c^1$  & \\
    $c$&   4& $(1,l_c^1)\times (-1,l_c^2)\times (-1,0)$ & $l_c^1$ & $-l_c^1$  &- &- &- &-  &0\\
     &    &  & $-l_c^2$  &$+l_c^2$ &  &  &  &  & \\
\hline
    4&$N^{(4)}$ & $(0,1)\times (0,-2)\times (1,0)$ & \multicolumn{7}{c|}{$x_A=4x_B=4l_c^2 x_C/l_c^1,\;\; l_c^1l_c^2=4-N^{(4)}/2$}\\
\hline
\end{tabular}%
\vskip 5.0in


\end{document}